\documentclass[floatfix,aps,twocolumn,prxquantum,superscriptaddress]{revtex4-2}
\usepackage{graphicx}% Include figure files
\usepackage{dcolumn}% Align table columns on decimal point
\usepackage{bm}% bold math
\usepackage[T1]{fontenc}
\usepackage{textcomp}
\usepackage{amsmath}
\usepackage{mathrsfs}
\usepackage{graphicx}
\usepackage{amssymb}
\usepackage{units}
\usepackage{xcolor}
\usepackage{float}

\newcommand{\plasmonium}{$\Pi\Pi$ qutrit}
\newcommand{\fluxonium}{$\Phi\Phi$ qutrit}
\newcommand{\plasmofluxonium}{$\Pi\Phi$ qutrit}
\newcommand{\fluxoplasmonium}{$\Phi\Pi$ qutrit}

\begin{document}

\title{Quantum Simulation with Fluxonium Qutrit Arrays}

\author{Ivan Amelio}
\email{ivan.amelio@ulb.be}
\affiliation{CENOLI,
Universit\'e Libre de Bruxelles, CP 231, Campus Plaine, B-1050 Brussels, Belgium}
\affiliation{International Solvay Institutes, 1050 Brussels, Belgium}

\author{Quentin Ficheux}
\affiliation{Univ. Grenoble Alpes, CNRS, Grenoble INP, Institut Néel, Grenoble, France}

\author{Nathan Goldman}
\affiliation{CENOLI,
Universit\'e Libre de Bruxelles, CP 231, Campus Plaine, B-1050 Brussels, Belgium}
\affiliation{International Solvay Institutes, 1050 Brussels, Belgium}
\affiliation{Laboratoire Kastler Brossel, Coll\`ege de France, CNRS, ENS-Universit\'e PSL, Sorbonne Universit\'e, 11 Place Marcelin Berthelot, 75005 Paris, France
}

\begin{abstract}
Fluxonium superconducting circuits were originally proposed to realize highly coherent qubits. In this work, we explore how these circuits can be used to implement and harness qutrits, by tuning their energy levels and matrix elements via an external flux bias. In particular, we investigate the distinctive features of arrays of fluxonium qutrits, and their potential for the quantum simulation of exotic quantum matter. We identify four different operational regimes, classified according to the plasmon-like versus fluxon-like nature of the qutrit excitations. Highly tunable on-site interactions are complemented by correlated single-particle hopping, pair hopping and non-local interactions, which naturally emerge and have different weights in the four regimes.  Dispersive corrections and decoherence are also analyzed. We investigate the rich ground-state phase diagram of  qutrit arrays and propose practical dynamical experiments to probe the different regimes. Altogether, fluxonium qutrit arrays emerge as a versatile and experimentally accessible platform to explore strongly correlated bosonic matter beyond the Bose–Hubbard paradigm, and with a potential  toward 
simulating lattice gauge theories and
non-Abelian topological states.
\end{abstract}

\date{\today}

\maketitle

\section{Introduction}

During the last years, the improvement of quantum hardware has reached the critical threshold  at which quantum devices  can perform computational tasks hardly achievable by classical computers~\cite{Arute2019,Daley2022,King2025}.
Superconducting circuits can be listed among the most mature quantum technologies, with  transmon qubits  being, at the moment,   a key of this success.
The transmon~\cite{Koch2007} consists of a capacitatively shunted Josephson junction (JJ), with a high Josephson energy ensuring excellent coherence times while retaining a sizable nonlinearity.
Nevertheless, alternative qubit architectures have been proposed and are being actively investigated, including fluxonium qubits~\cite{Manucharyan2009,Lin2018Demonstration,Nguyen2019,Moskalenko2022,Alibaba2022,Somoroff2023}, which consist of an inductively shunted transmon and enable to combine protection to charge noise with a larger anharmonicity.

Qubits based on superconducting circuits are very promising for realizing  fault tolerant universal digital quantum computers~\cite{Aharonov2008} in the long term.
However,  recent experiments 
have demonstrated that
tailored quantum simulators can already address important open problems of quantum physics, such as the phase diagram and the dynamics of quantum magnets~\cite{King2025} 
and of the Fermi-Hubbard model~\cite{alam2025programmabledigitalquantumsimulation}.
It is therefore strategic not to limit ourselves to the search for better qubits,
but to investigate also more general architectures for performing quantum simulations; in particular, it is believed that  {\em qudit}-based simulators can provide practical advantage in simulating lattice gauge theories and non-Abelian topological states~\cite{Mazza2010,Hafezi2014Engineering,Cuadra2022,Ciavarella2024,Meth2025}.

In the present paper,
we  propose to use state-of-the-art superconducting circuits  to engineer exotic
lattice Hamiltonians for bosons with three-body hard-core interactions.
More precisely,
we will consider an array of capacitatively or inductively coupled artificial atoms.
Each artificial atom provides three accessible states, the other levels being very off-resonant, and defines a {\em qutrit}.
In practice, the qutrits can be realized in fluxonium circuits where the JJ is inductively shunted by dissipation-free superinductors~\cite{Masluk2012}. The resonance condition between the levels of the qutrit can be achieved by adjusting the external magnetic flux. 
We will discuss this setup in details below, including estimates of their sensitivity to charge and flux noise.

The coupling between the qutrits results in unconventional processes for the photons excited in the system, such as correlated single-particle hopping, pair-hopping and two-body, three-body and four-body nearest-neighbor interactions. Moreover, 
the detuning between the levels of the qutrit provides a highly tunable Hubbard-like on-site interaction;
in contrast, transmons are typically limited to strongly attractive interactions.
The relevance of the different processes
can be broadly tuned by operating the qutrit in four different regimes, which we classify based on the nature  of the $0 \leftrightarrow 1$ and $1 \leftrightarrow 2$ transitions. Indeed, a transition can be plasmon-like,    involving a charge oscillation around a minimum of the Josephson potential, or fluxon-like, connecting two minima through a $2\pi$ phase-slip~\cite{Manucharyan2009,koch2009charging}.
We then refer to the four regimes as plasmon-plasmon ($\Pi\Pi$),
fluxon-fluxon ($\Phi\Phi$),
plasmon-fluxon ($\Pi\Phi$),
fluxon-plasmon ($\Phi\Pi$)
qutrits.

After deriving the general many-body Hamiltonian of the qutrit simulator, we discuss the phase-diagram of the ground state in a few characteristic regimes, mainly relying on the Gutzwiller approximation~\cite{Krauth1992,Sheshadri1993}.
Conventional  superfluids and Mott-insulators are predicted, together with more exotic pair superfluid, pair checkerboard and clustered states. We validate the mean-field predictions using exact diagonalization.
Moreover, we design simple experimental protocols where the dynamics brings characteristic hallmarks for each regime, without the need to cool the system to the ground state.

Similar many-body Hamiltonians arise in systems with flat bands~\cite{Huber2010,Takayoshi2013,Burgher2025}
and in frustrated quantum magnets~\cite{Zhitomirsky2010,Riberolles2024}.
The unconventional hopping and interaction terms also
appear when ultracold atoms are trapped in an optical lattice, but typically they provide small corrections on top of the standard Bose-Hubbard model arising in the harmonic limit of the trapping~\cite{Dutta2015}. 
Floquet engineering has been proposed as another route to implement these unconventional terms~\cite{Goldman2023}.
The interplay of pair-hopping, (extended) Hubbard interactions and the three-body constraint has been only partially explored and can give rise to very nontrivial physics and rich phase diagrams~\cite{Diehl2010,Mazza2010,Bonnes2011,Chen2011,Wang2013,Jorgensen2015,Malakar2023}.
Moreover, qutrit quantum simulators  have been previously  proposed  in the context of both ultracold atoms~\cite{Buchler2007,Daley2009,Mazza2010,Daley2014} and superconducting circuits \cite{Hafezi2014Engineering}. 
However, in these previous proposals the nearest neighbor interactions or correlated hopping terms are missing and the discussion is restricted to specific regimes.

Another fundamental application of qutrits would be the realization of the non-Abelian topological state known as the Pfaffian state~\cite{Moore1991}.
Indeed, the bosonic Pfaffian  is the  ground-state of a two-dimensional system of bosons in a (synthetic) magnetic field and with three-body hard-core repulsion, at unit filling of the lowest Landau level~\cite{Greiter1991,Mazza2010}.
In fact, the implementation of qutrit arrays in fluxonium circuits had already been proposed by Hafezi et al.~\cite{Hafezi2014Engineering} in this Pfaffian context.
However,  they restricted their proposal to   qutrits in the $\Pi\Pi$ regime.
Moreover, they overlooked the interaction contributions present for the inductive coupling.
We therefore aim to provide a more comprehensive and systematic study of qutrit quantum simulators.

The paper is structured as follows. In Section \ref{sec:general} we explain the general framework to map circuit Hamiltonians into coupled qutrits and highlight the origin of the unconventional hopping and interaction terms.
In Section \ref{sec:circuitQED}, we focus on the implementation in fluxonium circuits and analyze the four operational regimes.
In Section \ref{sec_many-body} we study both the ground state phase diagram and dynamics of the qutrit quantum simulators.
Finally, we draw our conclusions and outline future research directions in Section \ref{sec:outlook}.

\section{Qutrit quantum simulators}
\label{sec:general}

\begin{figure}[t]
    \centering
    \includegraphics[width=0.98\linewidth]{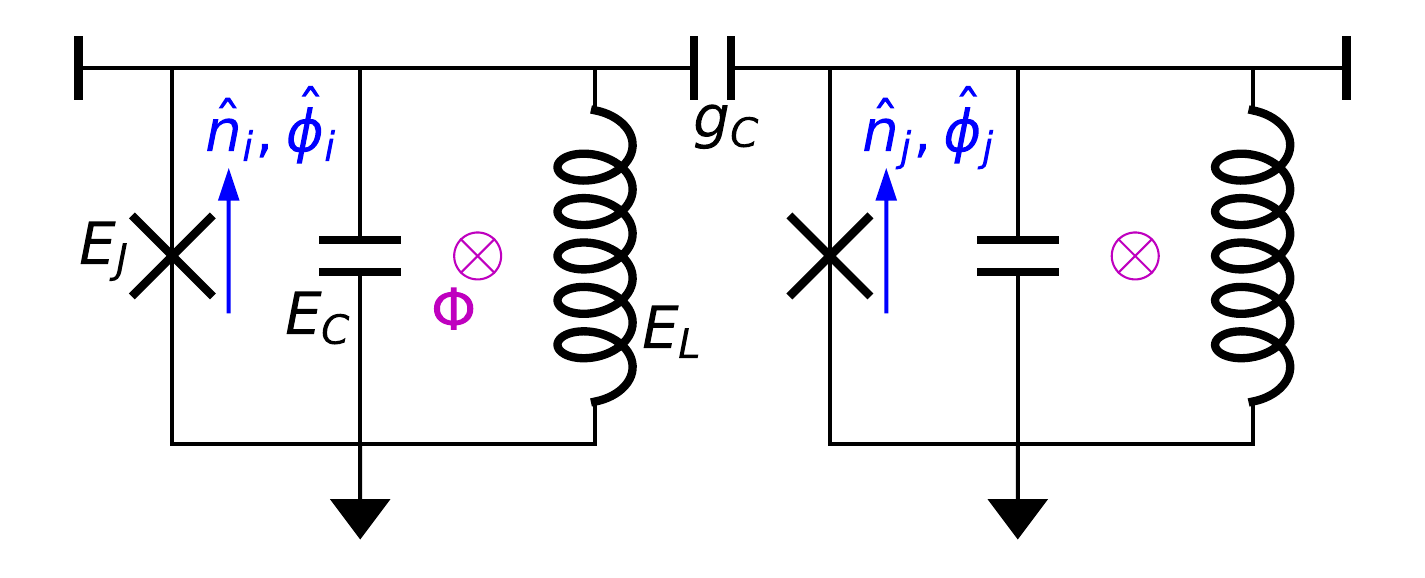}
    \caption{Sketch of two fluxonium circuits $i$ and $j$ capacitatively coupled.
    The operators $\hat{n}_j, \hat{\phi}_j$ describe the number and superconducting phase of the Cooper pairs with respect to the upper island of the circuit.
    The main control parameter is provided
    by the external magnetic  flux $\Phi$.
    }
    \label{fig:sketch}
\end{figure}

We consider a system composed of $L$ artificial atoms. The atom $j$  consists of a simple superconducting circuit described by the conjugate quantum degrees of freedom $(\hat{\phi}_j, \hat{n}_j)$, denoting the phase and Cooper pair number operators of the superconducting island, respectively.
An example with two capacitatively coupled fluxonium circuits is sketched in Fig. \ref{fig:sketch}.
In this work, we assume that the different atoms are only weakly coupled to each other. It is then convenient to  
introduce the Hamiltonian $\hat{H}_{j}$ of the $j$-th atom, and its spectral decomposition
\begin{equation}
    \hat{H}^{j} = \sum_{\bar{a}} \omega^j_{\bar{a}} \hat{\sigma}^j_{\bar{a}\bar{a}},
\end{equation}
where $\omega^j_{\bar{a}}$ is the eigenenergy of state $|\bar{a}\rangle_j$ (in units where $\hbar=1$) and
we introduced the notation $\hat{\sigma}^j_{\bar{a}\bar{b}}=|\bar{a}\rangle_j \langle \bar{b}|_j$. 
We use the upper bar notation to stress that these indices run over the full Hilbert space of each artificial atom, as opposed to the qutrit subspace introduced below.
The detailed circuit structure of an individual artificial atom and  the microscopic form of its Hamiltonian will be the topic of the next Section.

We will now discuss how the atoms are coupled to each other.
Assuming pairwise coupling, the Hamiltonian of the system takes the form
$ \hat{H} = \sum_j \hat{H}^{j} + \sum_{i<j} \hat{H}^{ij}$.
We consider here the two most common contributions, given by  capacitative and inductive couplings. 
Thus, the interaction Hamiltonian $\hat{H}^{ij}$ between atoms $i,j$ is given by the sum of the capacitative contribution
$\hat{H}_C^{ij} = g^{ij}_C \hat{n}_i \hat{n}_j$ and 
$\hat{H}_L^{ij} = g^{ij}_L \hat{\phi}_i \hat{\phi}_j$ for the inductive one.
The constants $g^{ij}_C$ and $g^{ij}_L$ are essentially determined by the (inverse) mutual capacitance and  inductance between the circuits, respectively.
Notice that, since in this work we only consider static couplings, it is convenient to absorb into the definition of the $\hat{H}^j$'s the corrections to the self-capacitance and self-inductance arising from the coupling elements.
In Appendix A, we demonstrate that, through small modifications in the design of the circuit,  both positive and negative $g^{ij}_C$'s and $g^{ij}_L$'s can be implemented.
Then, for the capacitative coupling contribution, we use the local basis and obtain the explicit expression
\begin{equation}
    \hat{H}_C^{ij}  =  
    \sum_{\bar{a} \bar{b} \bar{c} \bar{d}} 
    g_C^{ij} 
    n^i_{\bar{a} \bar{b}} 
    n^j_{\bar{c} \bar{d}} 
    \hat{\sigma}^i_{\bar{a} \bar{b}} 
    \hat{\sigma}^j_{\bar{c} \bar{d}},
\end{equation}
where 
$n^i_{\bar{a} \bar{b}} = 
\langle \bar{a}  | \hat{n}_i | \bar{b} \rangle
$;
 an analogous expansion holds for $H_L^{ij}$.
While this formalism is extremely general, we will now discuss several approximations relevant for qutrit quantum simulators.

First, we will assume that all the artificial atoms are identical and we will drop the dependence on $j$ of the local energy levels and of other matrix elements, such as
$n_{\bar{a} \bar{b}} = \langle \bar{a}  | \hat{n} | \bar{b} \rangle$ and 
$\phi_{\bar{a} \bar{b}} = \langle \bar{a} | \hat{\phi} | \bar{b} \rangle$.
The analysis of disorder due to fabrication imperfections goes beyond the scope of this work,
where we assume that the energy scale associated with disorder is much smaller than the coupling strength between the qutrits.

Moreover, connectivity is typically limited by the planar geometry of the devices. 
We thus assume that only some subset $\mathcal{B}$ of pairs of atoms are connected and all their coupling constants are identical.
We further neglect the weak longer range capacitative couplings that can arise in the Hamiltonian (see Appendix A)
and set  $g^{ij}_C=g_C$ and $g^{ij}_L=g_L$ if $(i,j) \in \mathcal{B}$, while $g^{ij}_C=g^{ij}_L=0$
otherwise. 
In the summations below,  we will then use the notation $<i,j>$ to sum over the indices of linked atoms.

We will now restrict the dynamics to three levels for each artificial atom, which will be thus called  {\em qutrit} in the rest of the paper. 
We select three levels,  that we call $|0\rangle,|1\rangle$ and $|2\rangle$, and we introduce the notation $\omega_{ab} = \omega_a -\omega_b$ for the energy splittings.
Notice that we drop the bar on the qutrit indices $a,b \in \{0,1,2\}$.
While we take $|0\rangle$ to be the ground state of the qutrit (i.e. $|0\rangle = |\bar{0}\rangle$), we do {\em not} require $|0\rangle,|1\rangle$ and $|2\rangle$ to be the three lowest states (e.g. $|2\rangle=|\bar{2}\rangle$ or $|2\rangle=|\bar{3}\rangle$ below).
We also define the detuning $\Delta = \omega_{21} - \omega_{10}$ and call $g \sim g_C,g_L$ the characteristic energy scale of the coupling Hamiltonian. 
Finally, we introduce  the detunings $\delta_{\bar{a}b}=||\omega_{\bar{a}b}| -\omega_{10}|$, for any combination of $\bar{a} \notin \{0,1,2\},b \in \{0,1,2\}$, and we denote as $\delta$ the smallest of such detunings.

Our goal is to emulate the dynamics of particles
which can hop between the different artificial atoms while conserving the particle number. Such particles correspond to the microwave photons excited in the system.
The conditions which we require for each  qutrit are then that: 
(1) the transitions $0 \leftrightarrow 1$ and $1 \leftrightarrow 2$ be quasi-resonant, or, in other words,
$|\Delta| \lesssim g$, the resonant condition corresponding to 
$\omega_{21} = \omega_{10}$;
(2) all the other levels be far off-resonant, i.e. $\delta \gg g$. 
Assuming that conditions (1) and (2) are met, we can truncate the local Hilbert space of each atom to the qutrit levels.
Then, it is also convenient to interpret the index $a$ as the number of photons at a given site.

With the extra assumption of large photon energy $\omega_{10}  \gg g$ (in practice this is often the case if condition (2) is met), we can also adopt the rotating wave approximation (RWA). The RWA consists in neglecting all  transitions which do not conserve the number of  excitations in the system, defined by the operator
$\hat{N} = \sum_j \hat{\rho}_j$, with $\hat{\rho}_j = \sum_{r=0,1,2} r \hat{\sigma}_{rr}^j$.
Within the RWA, it is very convenient to perform the transformation
$\hat{U}(t) 
=
\exp(i\omega_{10}\hat{N}t)$ 
to the frame where levels $0$ and $1$ have the same energy.
The resulting Hamiltonian reads
\begin{equation}
   \hat{H} =
   \Delta \sum_{j} \hat{\sigma}^j_{22} + \sum_{<i,j>} \sum_{abcd} g_{ab,cd} \hat{\sigma}^i_{ab} \hat{\sigma}^j_{cd},
   \label{eq:Hsigma}
\end{equation}
where $g_{ab,cd} = g_C n_{ab} n_{cd} +  g_L \phi_{ab} \phi_{cd}$ and the last sum runs over $a,b,c,d \in \{0,1,2\}$ with the constraint
$a+c=b+d$. This constraint implements the conservation of the particle number.

Finally, we assume that the qutrits are either capacitatively or inductively coupled, but not both. This assumption is experimentally reasonable and does not affect the theoretical framework, but allows to write some interaction terms in a more elegant way.
To this end, we introduce the operator
$\hat{b}_j^\dagger = \hat{\sigma}^j_{10} + \sqrt{2} \ \hat{\sigma}^j_{21}$~\footnote{
Without loss of generality, we make the gauge choice in the phase of the qutrit levels such that $n_{01},n_{12}>0$.
},
which describes the creation of a boson at site $j$ with the local Hilbert space truncation condition $\hat{\rho}_j=\hat{b}_j^\dagger \hat{b}_j \leq 2,
\ \forall j$, a condition also known as the 3-body hard-core constraint. 
The Hamiltonian is then recast into the equivalent representation
\begin{multline}
    H = - J \sum_{<i,j>} \alpha^{\hat{\rho}_i+\hat{\rho}_j-1} [\hat{b}^\dagger_i \hat{b}_j + h.c.]  +
    \frac{\Delta}{2} \sum_j  (\hat{b}^\dagger_j)^2 \hat{b}^2_j
    \\
    - \frac{P}{2} \sum_{<i,j>} [(\hat{b}^\dagger_i)^2 \hat{b}^2_j + h.c.]
    + \sum_{<i,j>} \hat{W}(\hat{\rho}_i,\hat{\rho}_j),
    \label{eq:the_H}
\end{multline}
which corresponds to an extended Bose-Hubbard model for photons with a set of exotic terms.
The first term describes {\em correlated} single particle hopping 
with rate $J=-g_C |n_{10}|^2$
and correlation coefficient  $\alpha = |n_{21}/\sqrt{2}n_{10}|$. When $\alpha=1$, one recovers  standard single particle hopping; notice that the factor $\sqrt{2}$ originates from bosonic enhancement.
The second term arises from the detuning of level $|2\rangle$ and, in the bosonic picture, corresponds to an onsite Hubbard interaction of strength $\Delta$.
The third term describes pair-hopping of photons with rate
$P=-g_C |n_{20}|^2$.
Notice that $J$ and $P$ can be negative or positive, but they have the same sign;
however, as explained in Appendix A, the sign of $J$ on bipartite lattices is inessential.
Similar expressions hold for inductive coupling.

Last but not least, we have a term diagonal in the qutrit level representation, giving rise to non-local interactions $\hat{W}(\hat{\rho}_i,\hat{\rho}_j)
=
\sum_{ab} g_{aa,bb} \hat{\sigma}^i_{aa} \hat{\sigma}^i_{bb}
$ between the photons.
Within the RWA, one has
$g_{aa,bb}= g_C n_{aa} n_{bb} +  g_L \phi_{aa} \phi_{bb}$.
In the qutrit implementations proposed below, we will always have $n_{aa}=0$,  the eigenfunctions of $\hat{H}^j$ being real in the phase representation. 
At the RWA level,
the $\hat{W}$ term is therefore  present only for inductively coupled qutrits, but non-resonant processes can generate $\hat{W}$ perturbative corrections, see Appendix B).
The non-local interaction term  can thus be factorized as 
$
 \hat{W}(\hat{\rho}_i,\hat{\rho}_j)
 = \hat{W}(\hat{\rho}_i) \hat{W}(\hat{\rho}_j)$,
 where
 $
 \hat{W}(\hat{\rho}) = \sum_r W_r |r\rangle \langle r |
 $ and  $W_r = \sqrt{g_L} \ \phi_{rr}$,
 which are imaginary numbers if $g_L<0$.
Physically, a finite $\phi_{rr}$ corresponds to a persistent DC  current in the circuit.
Rewriting this interaction as 
$ \hat{W}(\hat{\rho})
=
W_0 + (W_1 - W_0) \hat{\rho} 
+
\frac{1}{2}  (W_2 + W_0 - 2W_1) 
(\hat{\rho}-1) \hat{\rho}
= 
W_0 + \delta W_1 |1\rangle \langle 1 |
+ \delta W_2 |2\rangle \langle 2 |$, with $\delta W_r =W_r - W_0$,
makes it evident that $W_0$ only leads to a renormalization of the overall energy, of the chemical potential and of the on-site interaction $\Delta$. We thus focus on
$\delta W_1$ and $\delta W_2$ below, which encode nearest-neighbor interactions between particles and pairs, respectively.
 
When the ratio $g/\omega_{10}$ is small but finite, the parameters $J,\alpha,P,W_r$ are renormalized by non-resonant photon processes,  involving also  levels outside of the qutrit subspace. These corrections are studied in Appendix B, using both second order perturbation theory and exact diagonalization.
In other words, the Hamiltonian of Eq. (\ref{eq:Hsigma}) provides a more general framework than Eq. (\ref{eq:the_H}), with the couplings $g_{ab,cd}$  requiring perturbation theory in order to be precisely calculated.
(Notice that, even when the RWA is not  used to obtain the parameters in Eq. (\ref{eq:Hsigma}), this Hamiltonian still conserves the number of particles.)
To highlight the qualitative behavior of the qutrit simulator in the different regimes, below we will focus on  Eq. (\ref{eq:the_H})
and make use of the zeroth order RWA expressions for $J,\alpha,P,W_r$.

The bosonic Hamiltonian of Eq. (\ref{eq:the_H}), together with the local Hilbert space truncation, describes the qutrit quantum simulator, which is the main focus of this paper.
We will study its rich  many-body physics in Sec. \ref{sec_many-body}.
Before doing so, the next Section will be devoted to analyzing the implementation of qutrits  in fluxonium superconducting circuits.

\section{Qutrit realizations in fluxonium circuits}
\label{sec:circuitQED}

In the following, we will be considering arrays of artificial atoms built from superconducting circuits~\cite{2021Rasmussen}. The Hamiltonian of each atom is in the form
\begin{equation}
    H_{\rm at} = 4  E_C \hat{n}^2 + V(\hat{\phi}),
\end{equation}
where $\hat{n} = -i \frac{d}{d\hat{\phi}}$ is the charge operator and  
$E_C$ the charging energy. In the last term, 
$\hat{\phi}$ is the phase difference operator across the Josephson junction of the circuit and  $V(\hat{\phi})$ is the Josephson potential.

In conventional transmons, the cosine Josephson potential 
$V(\hat{\phi}) = - E_J \cos(\hat{\phi})$
entails that the energy needed for the second excitation is always red-detuned.
In a typical transmon, the ratio of the Josephson energy to the charging energy is of the order of $E_J/E_C \sim 40$-$100 $;  the detuning  
$\Delta \equiv \omega_{21} - \omega_{10} \sim -E_C$
is then in the range of a few hundred of MHz,
typically very large compared with $g$, the energy scale of the coupling between different artificial atoms. 
This situation is convenient for implementing two-level qubits, controllable with fast microwave pulses and highly coherent against charge noise.
One could reduce $\Delta/\omega_{10}$ by increasing $E_J/E_C$, but the increasing harmonicity of the spectrum poses a serious limit, since the circuit would behave as a linear oscillator~\cite{Viehmann2013}.
In a very recent work~\cite{ticea2025observationdisorderinducedsuperfluidity}, transmon-based qutrits enabled the observation of disorder-induced superfluidity; however, the largest reported hopping to Hubbard interaction ratio  was of the order of 1/10. 
In conclusion, transmons remain up to date a very promising architecture to achieve fault-tolerant digital quantum computation, yet their simple Hamiltonian restricts analog simulations to the regime of effectively hard-core photons.

Below, we will review and investigate fluxonium  circuits, where $\Delta \equiv \omega_{21} - \omega_{01}$ can be tuned across zero,  enabling weak and strong,  attractive and repulsive on-site interactions. For each configuration we also aim: 
(1) to verify that the other levels are far off resonance; 
(2) to characterize the matrix elements between the qutrit levels;
(3) to assess the experimental decoherence rates;
(4) to estimate up to which coupling scale $g$ the Hamiltonian of Eq. (\ref{eq:the_H}) is valid.
While fluxonium circuits seem the most convenient architecture to realize qutrits,
  in Appendix D we will also briefly consider realizations relying on higher-harmonic Josephson junctions.

\begin{figure*}[t]
    \centering
\includegraphics[width=0.48\linewidth]{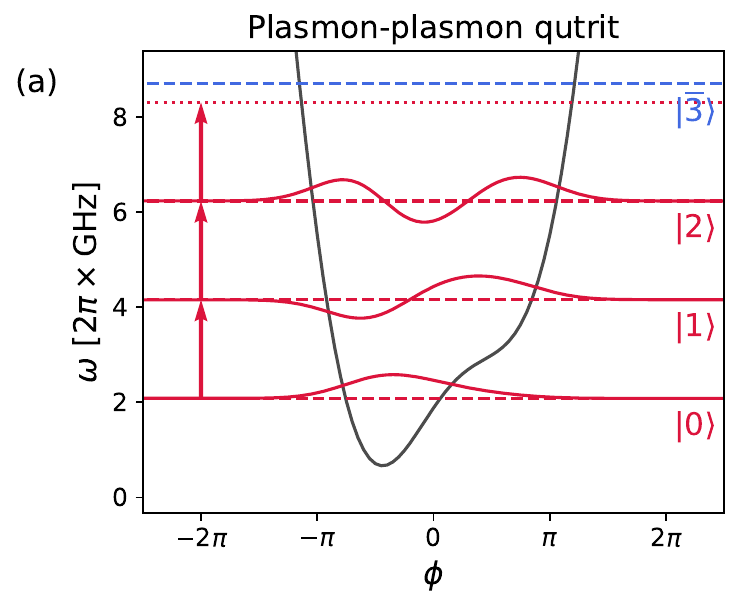}
\includegraphics[width=0.48\linewidth]{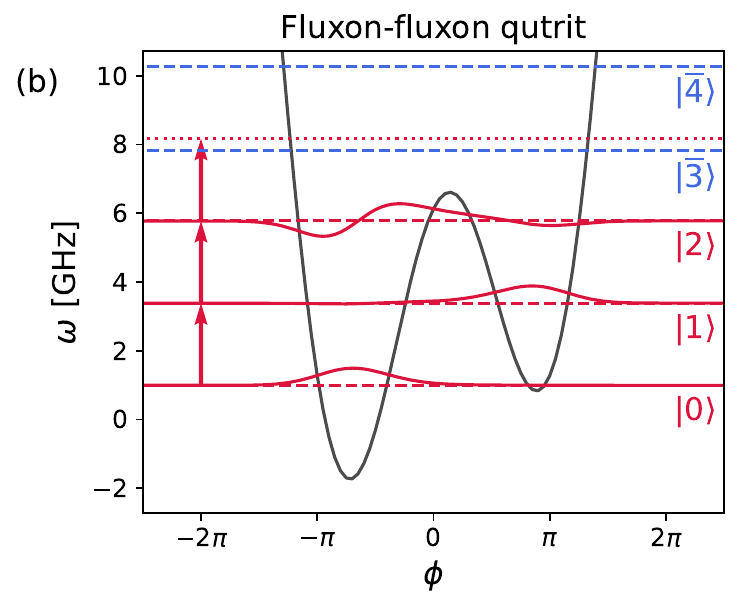}
\includegraphics[width=0.48\linewidth]{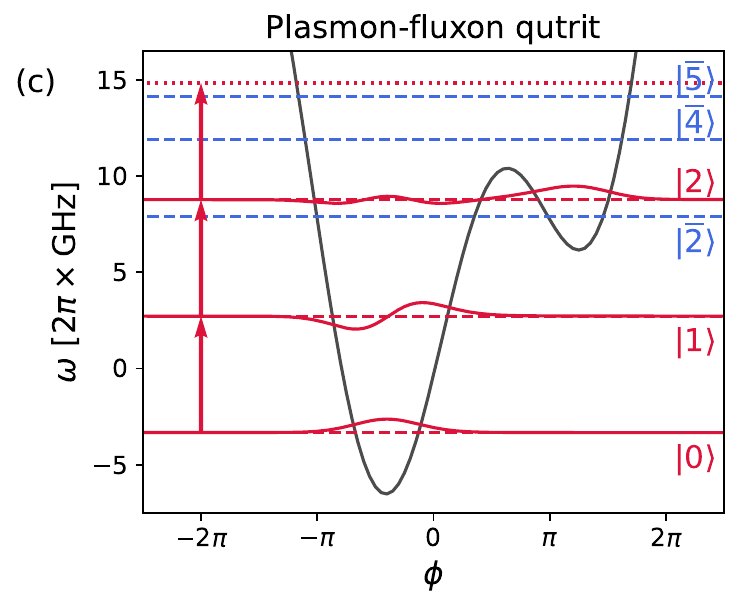}
\includegraphics[width=0.48\linewidth]{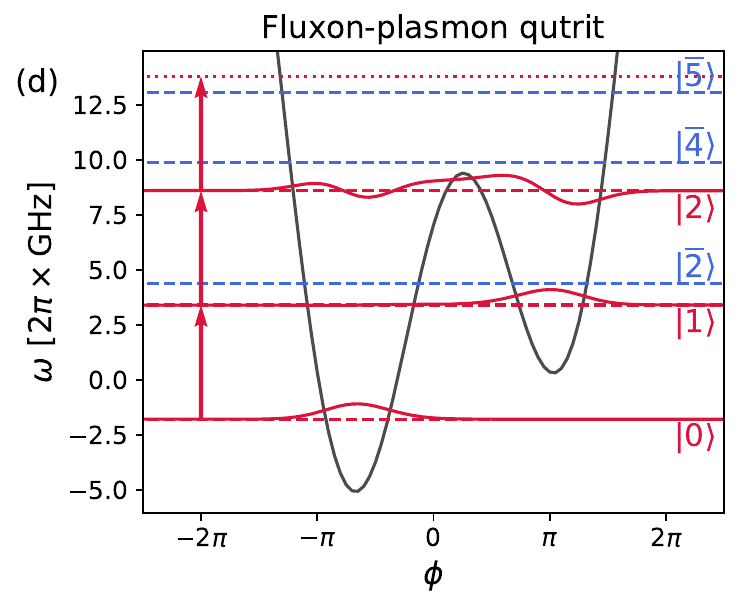}
    \caption{Qutrits arising in four different regimes of $H_{\rm at}$ are reported. The black
    solid line depicts the Josephson potential $V(\phi)$. 
    The red dashed
    lines represent the energy levels of the qutrit, and the solid red curves sketch the wavefunction of the states (not normalized). The blue dashed lines indicate the other levels
    of $H_{\rm at}$, which are off resonant with respect to the transitions at energy $\omega_{10}=\omega_{21}$ (in particular, the red dotted line provides a visual hint of the detuning). 
    The qutrit levels are labeled by their photon number $a=0,1,2$, while we used barred numbers to index the full spectrum of the circuit.
    Panels (a-d) correspond to the $\Pi\Pi$, $\Phi\Phi$, $\Pi\Phi$ and $\Phi\Pi$ qutrits. The names illustrate the nature of the  $0\leftrightarrow1$ and  $1\leftrightarrow 2$
    excitations of the qutrit.
    Notice that state $|2\rangle=|\bar{3}\rangle$ in the {\plasmofluxonium} and {\fluxoplasmonium}, since the second excited state of the circuit $|\bar{2}\rangle$ is off-resonant.
    }
    \label{fig:inductance_qutrits}
\end{figure*}

\subsubsection*{Fluxonium circuits}

The fluxonium circuit~\cite{Manucharyan2009} is shown in Fig. \ref{fig:sketch} and consists of a capacitor, a Josephson junction and a superinductance in parallel. In actual devices, the inductor element consists of either an array of large Josephson~\cite{Masluk2012} or hybrid~\cite{strickland2025gatemonium} junctions in series, a disordered superconductor~\cite{Hazard2019,grunhaupt2019granular}, or even a geometric superinductance~\cite{Peruzzo2020}. These architectures were used to implement qubits based either on fluxon transitions, which are protected from energy relaxation~\cite{Lin2018Demonstration,Earnest2018,Nguyen2019,Zhang2021,Somoroff2023,ardati2024using}, or plasmon transitions, yielding a more anharmonic version of the transmon~\cite{hassani2023inductively,Liu2023}.
Instead, the potential of fluxonium circuits to implement qutrits has only been partially studied in a theoretical work~\cite{Hafezi2014Engineering}, and experiments have primarily focused on using higher levels to control lower qubit states~\cite{Earnest2018,ficheux2021fast,Wang2024}.

The fluxonium circuit is described by the Hamiltonian
\begin{equation}
         H_{\rm at} = 4  E_C \hat{n}^2 - E_J\cos\left( \hat{\phi} + 2\pi\frac{\Phi ~}{\Phi_0} \right) + 
         \frac{E_L}{2}  \hat{\phi}^2,
    \end{equation}
where $\Phi$ describes the external magnetic flux threading the loop
through the junction and the superinductance, characterized by an inductive energy of $E_L$, and  $\Phi_0 = h/(2e)$ is the superconducting flux quantum.
Crucially, a finite $\Phi$ breaks the inversion symmetry of the Josephson potential and allows for $P,\delta W_r$ to be non-zero.
In experiments, the flux $\Phi$ is the easiest parameter to tune in-situ, and $E_J$ can also be adjusted if the Josephson junction is replaced by a SQUID loop with two junctions~\cite{Koch2007,Lin2018Demonstration,Alibaba2023}.

Importantly, the presence of the superinductor entails that the phase variable is now unwound and lives in $(-\infty, \infty)$ rather than being periodic on $[0,2\pi)$. As a result, a DC charge bias can always be gauged away,   suppressing dephasing from charge noise. Physically, this occurs because any DC charge accumulation would flow through the superinductance.  In contrast, the main source of decoherence in fluxonium circuits  originates from dielectric losses and external magnetic flux fluctuations~\cite{Nguyen2019,Alibaba2023}. To achieve the resonance condition required in this work, we need to operate the circuits away from their flux-insensitive sweet spot $\Phi=0.5\Phi_0$. We will therefore provide for each qutrit an estimate of the sensitivity to flux noise.

In the following, we will separately analyze four different sets of parameters, which correspond to characteristic level structures of the qutrits and allow to realize four different regimes for the effective constants $\alpha$ and $P/J$ appearing in the general Hamiltonian of Eq. (\ref{eq:the_H}).
Notice that we will fix $E_C = 2 \pi \times 0.60$~GHz and $E_L = 2 \pi \times 1.50$~GHz, while we vary $E_J$. This shows that the four regimes can be achieved with a single device, since $E_J$ can be adjusted via an external flux if the Josephson junction is replaced with a SQUID.
As an example of parameters used in current devices, in the experiment by Liu et al. \cite{Liu2023} the superinductance was made of 37 Al/$\mathrm{AlO_x}$/Al JJs in series and their device had $E_C = 2 \pi \times 0.60$~GHz, $E_J = 2 \pi \times 5.61$~GHz and $E_L = 2 \pi \times 2.20$~GHz. 
We refer to Table \ref{tab:4regimes}
for a summary of the results obtained in the four regimes.

\subsubsection*{Plasmon-Plasmon Qutrit}

\begin{table*}[t]
\centering
\begin{tabular}{ |c|c|c|c|c|c|c|c|c|c|c| }
\hline
Qutrit regime & 
$E_J$ [$2\pi\times$GHz] & $\Phi_{\rm ext}/\Phi_0$ &
$\omega_{10}$ [$2\pi\times$GHz] & $\delta$ [$2\pi\times$GHz] &
\ \ $\alpha$ \ \ & $P / J$ &
\ \ \ $w_1$ \ \ \ & \ \ \ $w_2$ \ \ \ &
\ $T^{\rm diel}_{ab}$ [$\mu$s] \ & \ $T^{\Phi}_{ab}$ [$\mu$s] \
\\
 \hline
 Plasmon-plasmon & 
2.2 & 0.413 &
2.08 & 0.393 &
1.03 & 0.29 &
 1.40 & 1.22 &
24.5$|_{1 \leftrightarrow 2}$ & 13.9$|_{0 \leftrightarrow 1}$ \\ 
 Fluxon-fluxon  & 
6.5 & 0.446 &
2.39 & 0.356 &
2.8 & 131 &
 48.2 & 1.4 &
21.8$|_{0 \leftrightarrow 2}$ &
3.6$|_{0 \leftrightarrow 1}$ \\ 
 Plasmon-fluxon  & 
8.0 & 0.243 &
6.06 & 0.697 &
0.28 & 4$\cdot 10^{-4}$ &
0.05 & 50.4 &
13.2$|_{0 \leftrightarrow 1}$ & 
3.8$|_{0 \leftrightarrow 2}$ \\  
 Fluxon-plasmon  & 
9.0 & 0.393 &
5.20 & 0.750 &
12.1 & 7.8 &
8612 & 4151 &
23.8$|_{1 \leftrightarrow 2}$ & 
3.3$|_{0 \leftrightarrow 1}$ \\  
 \hline
\end{tabular}
\caption{
In this table we characterize the four qutrit regimes obtainable in a fluxonium circuit with $E_C=2\pi\times 0.60$~GHz and $E_L=2\pi\times 1.50$~GHz.
From left to right, we report in each column the Josephson energy $E_J$, the external flux bias $\Phi/\Phi_0$, the qutrit transition frequency $\omega_{10}$, the smallest detuning $\delta$ of the additional fluxonium levels from the qutrit levels, the hopping correlation $\alpha$,
the pair hopping over single particle hopping rate $P/J$, the coefficients of the non-local interactions $w_r = -\delta W_r^2 / (J+P)$, the maximum achievable lifetime due to dielectric losses
$T^{\rm diel}_{ab}$
and the dephasing time caused by flux noise  $T^{\Phi}_{ab}$ (we indicate also the least coherent transition as subscript).
We report here $P/J$ calculated for capacitative coupling, which is four times the one from the inductive coupling case.
All energies and frequencies are in units of $2\pi\times$GHz. 
}
\label{tab:4regimes}
\end{table*}

We call the first configuration the    plasmon-plasmon or {\plasmonium} and we illustrate it in Fig.~\ref{fig:inductance_qutrits}.(a). The  idea behind the ``plasmonium qubit'' introduced by Liu et al.~\cite{Liu2023} is based on the remark that, in the transmon, having a pure Josephson potential entails a tradeoff between  coherence and anharmonicity.
    In this regime, the Josephson potential is used to make the harmonic potential of the superinductance slightly asymmetric.
    This can be used to realize highly coherent qubits
    with larger anharmonicity
    than the transmon. 
    In our case, we can exploit the asymmetry to achieve the resonance condition $\omega_{10}=\omega_{21}$.
    Notice that already the authors of \cite{Hafezi2014Engineering} noticed that this configuration allowed to achieve the resonance condition (they were interested in the realization of Pfaffian states), but they limited themselves to the situation $\alpha \simeq 1$, i.e. to the Hubbard model with the three-boson hard core constraint. Also,
    they proposed to use inductive coupling so to have a negligible $P/J$ and to implement synthetic Berry phases, but they overlooked the presence of the $\hat{W}$ contributions, which are sizable for inductive coupling.

    As visible in Fig.~\ref{fig:inductance_qutrits}.a,
     the name {\plasmonium} reflects the excitations of the system being plasmonic. A plasmon excitation 
occurs around a minimum of the Josephson potential and involves an oscillation of energy between charge and phase, 
     with only a small average change of phase. 
For 
$E_J = 2 \pi \times 2.2$~GHz and
an external flux of $\Phi \simeq 0.413 ~\Phi_0$, we thus find
$\omega_{10} = \omega_{21} \simeq 2 \pi \times 2.08$~GHz, while the fourth level is  blue-detuned by $\delta =  2 \pi \times 393$~MHz.
For capacitative coupling, pair hopping is expected to be small but not negligible, $P/J|_C = 0.29$, with  $\alpha|_C \simeq 1$ corresponding to standard single particle hopping. Notice that the diagonal matrix elements of $\hat{n}$ are exactly zero, since the wavefunctions can be chosen real.
For inductive coupling, 
the exact identity 
$\omega_{ab} \phi_{ab}
=
-i8E_C n_{ab}
$
entails  $\alpha|_C=\alpha|_L$ and
$P/J|_C=4P/J|_L$. Thus, for brevity, in the following we will report only the capacitative values. 
The interaction term $\hat{W}$,
quantified via 
$w_r = -\delta W_r^2 / (J+P)$ (the choice of the denominator is motivated by the fact that in some regimes either $J$ or $P$ are very small),
is also sizable, resulting in $w_1=1.40, w_2=1.22$.
Notice that with this convention $w_1$ and $w_2$
are always positive at the RWA level, but the sign of the physical interaction $\delta W_r^2$ is given by the sign of $g_L$ (repulsive for $g_L>0$, and vice-versa). 
The coherence properties are discussed in an ad-hoc paragraph below; the effective qutrit parameters are summarized in Table \ref{tab:4regimes}.

\subsubsection*{Fluxon-Fluxon Qutrit}

The second regime that we consider will be called 
\fluxonium~and is illustrated in Fig.~\ref{fig:inductance_qutrits}.(b). The name is motivated by the fact that
 the transitions $0 \to 1$ and $1 \to 2$ correspond to the excitation of a ``fluxon'', involving a phase change of approximatively $2\pi$~\cite{koch2009charging}. The transition $0 \to 2$, in contrast, is of the plasmon type.
 The small matrix elements associated with fluxon transitions
 have been exploited to realize fluxonium qubits with
 very long-lived excitations~\cite{Lin2018Demonstration,hassani2023inductively}. Moreover, the strong departure from a parabolic-like potential provides strong nonlinearities.

These considerations apply also to the \fluxonium. 
In Fig.~\ref{fig:inductance_qutrits}.(b),
we take 
 $E_J = 2 \pi \times 6.5$~GHz
and achieve the resonance condition $\Delta=0$ by setting $\Phi \simeq 0.446 ~\Phi_0$.
For capacitive coupling, this yields an anomalous $\alpha \simeq 2.8$, but the dominant process is pair-hopping with $P \simeq 131 J$. 
The non-local interaction term features $w_2=1.4$ together with a gigantic $w_1=48.2$. This suggests that, in a qutrit array with $g_L>0$, clusters of one photon per site are strongly favorable in energy, while the physics will be dominated by pairs for $g_L<0$.

\subsubsection*{Plasmon-Fluxon Qutrit}

We term the third configuration \plasmofluxonium, since, as visible in Fig.~\ref{fig:inductance_qutrits}.(c), the transition $0 \to 1$ is of the plasmon type, but the $1 \to 2$ involves a large phase variation.
The \plasmofluxonium~can be realized with  $E_J=2 \pi \times 8.0$~GHz and by setting the flux to $\Phi \simeq 0.248 ~\Phi_0$. A main difference with respect to the two previous configurations  is that now 
the level $2$ of the qutrit is provided by the fourth level of the $\hat{H}_{\rm at}$, i.e. $|2\rangle = |\bar{3}\rangle$, while $|\bar{2}\rangle$ is red-detuned by 
$E_{\bar{3}} - E_{\bar{2}}
=
2 \pi \times 859$~MHz.
Also, in this case $\omega_{10}\simeq 2 \pi \times 6.06$~GHz
and the sixth energy level is off-resonant by $\delta =  2 \pi \times 697$~MHz, see the highest blue dashed line and the red dotted line in Fig.~\ref{fig:inductance_qutrits}.(c).
The specificity of this qutrit configuration is that $\alpha < 1, P \ll J$. More precisely, the chosen parameters yield $\alpha \simeq 0.28$ and $P/J\simeq 4\cdot 10^{-4}$ for capacitative coupling.
For inductive coupling, $w_1=0.05$ and $w_2=50.4$ entail strong nearest-neighbor interactions when pairs are present.

\begin{figure*}[t]
    \centering
\includegraphics[width=0.98\linewidth]{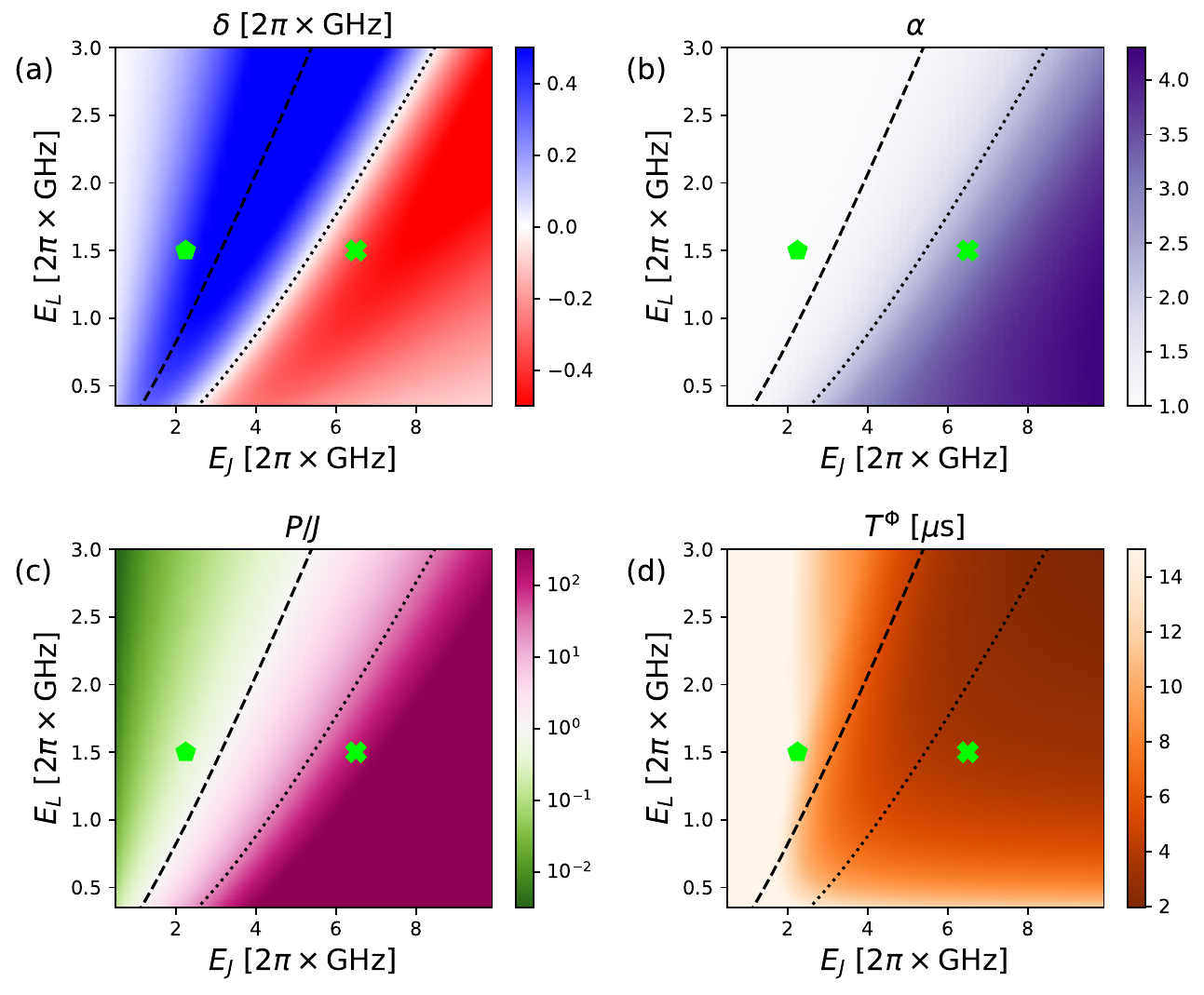}
    \caption{Qutrit parameters for different inductive energies $E_L$ and Josepshon junction energies $E_J$. In this plot, the qutrit levels 0,1,2 correspond to the three lowest eigenstates of $H_{\rm at}$. 
    In panel (a), we plot the detuning $\delta$ of the closest off-resonant level. The black dotted line indicates where $\delta$ crosses zero.
    Panel (b) reports the degree of correlation of single-particle hopping $\alpha$.
    Panel (c) illustrates the relevance of pair-hopping processes through $P/J$ (in log scale).
     For reference, the black dashes indicate the line where $P/J=1$. 
     Panel (d) 
     reports the dephasing time due to noise in the external flux bias.
     In all four panels, the light green pentagon corresponds to the \plasmonium~of Fig.~\ref{fig:inductance_qutrits}.(a), and the cross to
     the \fluxonium~of Fig.~\ref{fig:inductance_qutrits}.(b).}
    \label{fig:4panel_LJ}
\end{figure*}

\subsubsection*{Fluxon-Plasmon Qutrit}

The last configuration
will be denoted as \fluxoplasmonium.
As the name suggests, the nature of the transitions is reversed with respect to the \plasmofluxonium: 
the excitation $0 \to 1$ is basically a fluxon and  the $1 \to 2$ one is plasmonic. 
In Fig.~\ref{fig:inductance_qutrits}.(d)
we plot the {\fluxoplasmonium} level structure 
using as parameters  $E_J=2 \pi \times 9.0$~GHz and $\Phi=0.393 ~\Phi_0$. Also in this case $|2\rangle = |\bar{3}\rangle$.
We find that $\omega_{10}=2 \pi \times 5.20$~GHz and between the extra levels the least off-resonant one is $|\bar{5}\rangle$, red-detuned by $\delta = 
\omega_{\bar{5}} - \omega_{2} - \omega_{10}
=
-2 \pi \times 750$~MHz.
The peculiarity of the \fluxoplasmonium~is the fact that the 
$1 \leftrightarrow 2$
matrix element  is predominant, as quantified by $\alpha \simeq 12$ and $P/J \simeq 8$.
The very large values of $w_1$ and $w_2$ suggest
that the {\fluxoplasmonium}s should be coupled capacitively in order to display interesting many-body physics, avoiding being dominated by the nearest-neighbor interactions.
To our knowledge,
the potential of the \plasmofluxonium~and the \fluxoplasmonium~to generate exotic hopping processes has not been studied in the literature yet.

\subsubsection*{Coherence times}

In this section, we review the dominant energy relaxation and decoherence mechanisms affecting fluxonium qubits and estimate their associated time scales. These estimates define the typical duration over which the system can be regarded as effectively isolated from its environment and
can be accurately described only by Hamiltonian Eq.~(\ref{eq:the_H}).

Several previous experiments have derived and benchmarked noise models for fluxonium circuits~\cite{Nguyen2019,Somoroff2023,Alibaba2023,ardati2024using}. In most cases, the energy relaxation of fluxonium is dominated either by dielectric loss or by magnetic flux noise in the circuit loop. With our parameters, we expect dielectric loss to be the dominant factor~\cite{Alibaba2023}. The inverse energy relaxation time $1/T^{\rm diel}_{ab} = \Gamma^{\rm diel}_{ab}$ for the transition $a \leftrightarrow b$ is then given by Fermi's golden rule
\begin{equation}
    \Gamma^{\rm diel}_{ab}
    =
    \frac{\hbar \omega_{a b}^2}{4 E_C} |\phi_{ab}|^2 \tan\delta(\omega_{ab}) 
    \coth \left( \frac{\hbar\omega_{ab}}{2k_B T}
    \right),
    \label{eq:Tdepola}
\end{equation}
where the loss tangent $\tan\delta(\omega)$ has a weak frequency dependence and was fitted to 
$\tan\delta(\omega) \simeq 2 \cdot 10^{-6}
\left(
\frac{\omega}{2\pi \times 6\mathrm{GHz}}
\right)^{0.15}$ for a fluxonium qubit on a silicon substrate~\cite{Nguyen2019,Alibaba2023}.
The last term provides finite temperature corrections ($T$ being the temperature and $k_B$ the Boltzmann constant). In this work, we assume working at a temperature of 20~mK, while the $\omega_{ab}$'s are of the order of a few GHz, the hyperbolic cotangent is therefore close to 1.
We have calculated the depolarization times for the four qutrits of Fig.~\ref{fig:inductance_qutrits}.
The shortest lifetimes are obtained for plasmon-like transitions, where the charge matrix element is sizable, while fluxon transitions feature very small depolarization rates. The shortest depolarization time is found for 
the $0 \leftrightarrow 1$ transition of the $\Pi\Phi$ qutrit, and equals $T^{\rm diel}_{01} = 13 ~\mu$s. As expected, this is much shorter than the millisecond lifetimes achievable in fluxonium qubits~\cite{Somoroff2023}.
The plasmon-like transitions  display comparable $T^{\rm diel}$'s also in the other qutrit regimes.
Moreover, notice that in the $\Pi\Phi$ qutrit the depolarization rate between the extra level $\Bar{2}$ and qutrit level 1 
has $T^{\rm diel}_{1\Bar{2}} = 8.6 ~\mu$s. However, this will not affect the operation in the qutrit subspace, since the $2\to\Bar{2}$ transition is fluxon-like and has a small matrix element, while the thermal excitation $1\to\Bar{2}$ is strongly suppressed by its large energy cost $\sim 2\pi \times 5$~GHz.
Similar considerations protect  $\Phi\Pi$ qutrit from depolarization into level $\Bar{2}$.

The coherence of the qutrit is instead limited by dephasing induced by low-frequency fluctuations of the external flux bias away from the first-order insensitive flux sweet spots~\cite{Nguyen2019}. The fact that the power spectrum of flux noise is peaked at low frequency, with a noise spectral density close to $1/f$~\cite{ateshian2025temperature}, entails that in experiments the signal of Ramsey spectroscopy over time decays as a Gaussian~\cite{Ithier2005}.
In mathematical terms,
the off-diagonal elements of the density matrix $\hat{\xi}$ decay as
$\xi_{ab} \sim e^{-t/T^{\rm diel}_{ab}}
e^{-(t/T^{\Phi}_{ab})^2}$.
The dephasing time $T^{\Phi}_{ab}$ is related to the derivative of the 
$a \leftrightarrow b$
transition energy over flux bias, through the formula
\begin{equation}
    1/T^{\Phi}_{ab}
    =
    A \left|
\frac{\partial \omega_{ab}}{\partial \Phi}
    \right|,
    \label{eq:Tdeph}
\end{equation}
where values $A \sim 10^{-6} \Phi_0$ have been reported experimentally~\cite{Yoshihara2006,Nguyen2019,ateshian2025temperature}.

The most coherent fluxonium qubits~\cite{Somoroff2023} are operated at the sweet spot
$\Phi=0.5\Phi_0$, for which $\frac{\partial \omega_{10}}{\partial \Phi}=0$.
In this case, 
$T^{\Phi}_{10}$ needs to be evaluated at a higher order in perturbation theory and can be well above milliseconds. The coherence time is instead limited by energy relaxation~\cite{Nguyen2019} or by additional decoherence mechanisms, such as quasiparticle poisoning~\cite{pop2014coherent}, photon shot noise~\cite{sears2012photon}, or quantum phase slips~\cite{Randeria2024}, which can be mitigated through proper circuit design.
For the qutrits presented in Fig.~\ref{fig:inductance_qutrits}, the flux needs to be tuned away from the sweet spot in order to achieve the resonance condition $\Delta \simeq 0$.
Thus, flux noise provides the most limiting source of decoherence.  
For the parameters of Fig.~\ref{fig:inductance_qutrits}, the least coherent transition is the $0 \leftrightarrow 1$ one of the \fluxoplasmonium, with $T^{\Phi}_{10} \simeq 3.3~\mu$s.
Similar coherence times are obtained for the fluxon-like transitions, while the plasmon-like transitions are typically one order of magnitude more coherent.

The estimates provided here are based on previous fundamental studies of fluxonium circuits and we expect that the improvement and optimization of device fabrication can lead to an improvement of a few orders of magnitude.
We remark that, using the same values for 
$E_C$ and for the loss tangent,
a transmon with $E_J=40E_C$ would have a depolarization lifetime of 7.3~$\mu$s, on-par with the ones reported here. 
In summary, we expect that the qutrits based on fluxonium circuits can reach coherence times comparable to transmon qubits.

\subsubsection*{Parameter exploration}
These four examples suggest that the qutrit Hamiltonian of Eq.~(\ref{eq:the_H}) can be realized with fluxonium circuits and  very different regimes for the parameters $\{ \alpha,P/J,w_r \}$ can be achieved. Moreover, adjusting the flux allows one to sweep $\Delta$ through zero.
Having illustrated these four level structures with specific parameters, we now report in Fig.~\ref{fig:4panel_LJ} a more comprehensive characterization of
the behavior of the lowest three states of $H_{\rm at}$
as a function of $E_L$ and $E_J$.
For each ($E_L$, $E_J$) point, we fix $E_C=2 \pi \times 0.60$~GHz and
 tune $\Phi$ so to achieve the resonance condition 
 $\Delta=0$.
In this plot, we take as states  of the qutrit the three lowest levels of $H_{\rm at}$, i.e. $|0\rangle=|\bar{0}\rangle,|1\rangle=|\bar{1}\rangle,|2\rangle=|\bar{2}\rangle$. This choice captures the $\Pi\Pi$ and $\Phi\Phi$ qutrits described above; in particular,  the parameters of Fig. \ref{fig:inductance_qutrits}.(a) and (b) are indicated by the green pentagon and cross, respectively.
The excitation frequency $\omega_{10}/2\pi$ lies in the range between 1 and 4 GHz for this range of $E_L$'s and $E_J$'s.

In Fig. \ref{fig:4panel_LJ}.(a), we plot $\delta=E_{\bar{3}}-E_2-\omega_{10}$, the detuning of the fourth level. A blue and a red regions can be distinguished, where the fourth level is blue-detuned and red-detuned, respectively. In between, there is the black dotted line where 
the qutrit approximation $|\delta| \gg g$ is expected to break down and a 4-level atom emerges. When crossing this line from blue to red, we find that level $|2\rangle$ becomes localized in the same well as level $|0\rangle$.

In Fig. \ref{fig:4panel_LJ}.(b), we report $\alpha$,
the degree of correlation in the single particle hopping term of the qutrit array Hamiltonian.
It turns out that all over the $\Pi\Pi$ regime $\alpha$ is very close to 1, while in the $\Phi\Phi$ regime  $\alpha >1$. To achieve $\alpha<1$, one should instead resort to the  $\Pi\Phi$ regime.

In Fig. \ref{fig:4panel_LJ}.(c), we report $P/J$ (in logarithmic scale),
quantifying the importance of pair-hopping processes.
We use the black dashed line where $P/J=1$ to define the transition between the \plasmonium~and  \fluxonium~regimes.
In the $\Pi\Pi$ regime $P/J <1$, while in the $\Phi\Phi$ regime pair-hopping   represents the leading kinetic process.

In Fig. \ref{fig:4panel_LJ}.(d), we consider the effect of dephasing induced by the external magnetic flux fluctuations
and we plot $T^{\Phi}$ for the least coherent transition.
The coherence time displays a clear dependence on $E_J$ and for the  \plasmonium~can be one order of magnitude larger than the \fluxonium~for the parameters studied here.

For completeness, we also report in Appendix C the behavior of the qutrit parameters with $E_J$ when the qutrit is formed by the first, second and fourth levels of $H_{\rm at}$, i.e. $|0\rangle=|\bar{0}\rangle,|1\rangle=|\bar{1}\rangle,|2\rangle=|\bar{3}\rangle$. This situation includes the plasmon-fluxon and fluxon-plasmon qutrits.

In conclusion, by  operating fluxonium circuits in four very different regimes, we have demonstrated the possibility of realizing qutrit arrays described by the Hamiltonian 
of Eq. (\ref{eq:the_H}), with a broad tunability of the parameters $\{\alpha, P/J,\Delta,w_r\}$ and good coherence properties.
In Appendix D, we consider an alternative circuit architecture
to realize qutrits, based on higher-harmonic Josephson junctions. We show that similar Josephson potentials and level structures can be achieved, but very high decoherence rates are expected.

\section{Quantum many-body physics}
\label{sec_many-body}

We have so far established that the Hamiltonian $\hat{H}$ of Eq. (\ref{eq:the_H}) can be realized with superconducting circuit devices in many different regimes of parameters.
Thus, we can now focus on the quantum physics described by $\hat{H}$, independently of its microscopic realization.
Here, we will first give an overview of the rich phase diagram for the ground state of $\hat{H}$. 
Then, we will investigate the Hamiltonian dynamics starting from easily realizable product states.

\subsection{Ground-state phase diagram}

\begin{figure}[t]
    \centering
\includegraphics[width=0.98\linewidth]{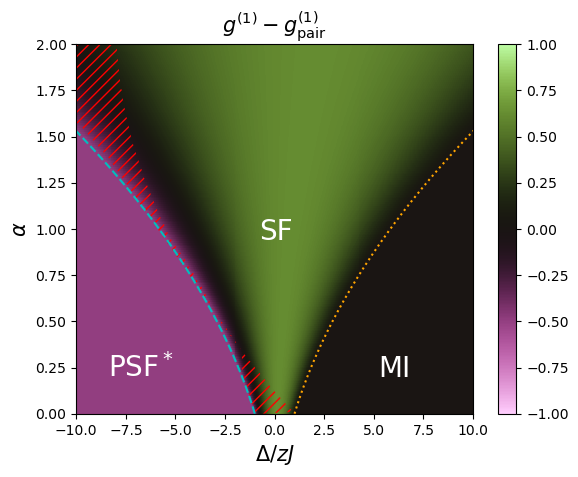}
    \caption{Difference between the single-particle coherence $g^{(1)}=
\langle b^\dagger_i \ b_j \rangle$
and the pair coherence $g_{\rm pair}^{(1)}=
\langle (b^\dagger_i)^2 \ b_j^2 \rangle$,
calculated for the ground-state of $H_{\alpha\Delta}$ at unit filling $n=1$, within the Gutzwiller approximation. We can distinguish a standard superfluid (SF), a  pair superfluid with small stiffness (PSF$^*$) and a Mott insulator phase (MI) phase.
Below the cyan dashed line we have $\psi_1=0$, while $\psi_0=\psi_2=0$ below the orange dots.
The red hatched area 
is defined by 
the thermodynamic instability condition
$\frac{d^2 e}{dn^2} <0$.
}
\label{fig:Gutzwiller_alfaDelta_fill1_main}
\end{figure}

In order to characterize the zero-temperature phase diagram of $\hat{H}$,
we are going to adopt the bosonic Gutzwiller approximation~\cite{Krauth1992,Sheshadri1993}.
The Gutzwiller ansatz consists in assuming that the wavefunction of the system is a tensor product of local wavefunctions.
This approach has been established as a simple yet effective description of the Mott and superfluid phases of the standard Bose-Hubbard model~\cite{Krauth1992,Sheshadri1993} as well as of exotic phases arising in extended Bose-Hubbard models, such as pair-superfluids, charge-density-waves and atomic or pair supersolids~\cite{Wang2013,Malakar2023}.
Our mean-field results are then validated in Appendix F, where we use exact diagonalization (ED) to compute  distinctive  correlation functions in the ground state. 

 The Gutzwiller approximation for the ground state $|\Psi\rangle$ of the array of qutrits reads
\begin{equation}
    |\Psi\rangle = \bigotimes_j \  (\psi_0^{(j)} |0\rangle_j + \psi_1^{(j)} |1\rangle_j+ \psi_2^{(j)} |2\rangle_j ),
\end{equation}
where the $\{\psi_a^{(j)} \}_{a=0,1,2}$ specify the local wavefunction of  qutrit $j$.

Here, we assume that both $J$ and $P$ are positive (as shown in Appendix A, the case $J<0$ is essentially equivalent) and the qutrits are arranged in a translationally invariant bipartite lattice, with $C$ and $D$ sublattices, and coordination number $z$~\footnote{In practice, we have in mind square lattices in $d=1$ or $d=2$ dimensions, for which $z=2d$.}.
We start considering translationally invariant states
$\psi_r^{(j)} = \psi_r$ (the sublattice symmetry will be broken below, where the $\hat{W}$-term is considered).
 A single-particle condensate is signaled by a finite value of the off-diagonal coherence
$g^{(1)}=
\langle \hat{b}^\dagger_i \hat{b}_j \rangle
=
|\psi_0^*\psi_1 + \sqrt{2}\psi_1^*\psi_2|^2$, with $i \neq j$. Similarly, condensation of pairs corresponds to $g_{\rm pair}^{(1)}=
\langle (\hat{b}^\dagger_i)^2 \ \hat{b}_j^2 \rangle
= 2|\psi^*_0\psi_2|^2$. 
At density $n=1$, a Mott insulating state can occur with $\psi_1=1, \psi_0=\psi_2=0$. 
Moreover, the thermodynamic stability 
of the state is assessed by calculating $\frac{d^2 e}{dn^2}$, with $e$ the expectation value of the Hamiltonian per site. When this is negative, it signals instabilities towards forming high density clusters.
In a closed system, this corresponds to phase separation.
For a system connected to a particle reservoir, the grand-canonical energy would display a first-order phase transition in the chemical potential.
A detailed description of the Gutzwiller calculations is reported in Appendix E.

\subsubsection{Phase diagram in the  $(\alpha,\Delta)$ plane}

We will start by considering the case where pair-hopping processes can be neglected and $W_r=0$,
so that the Hamiltonian reduces to
\begin{equation}
     H_{\alpha\Delta} = - J \sum_{<i,j>} \alpha^{\rho_i+\rho_j-1} [\hat{b}^\dagger_i \hat{b}_j + h.c.]  +
    \frac{\Delta}{2} \sum_j  (\hat{b}^\dagger_j)^2 \hat{b}^2_j.
\end{equation}
In our proposal, the small $\alpha$ regime can be implemented in the  \plasmofluxonium, while large $\alpha$ is achieved in the \fluxoplasmonium~scenario. Capacitative coupling ensures
$W_r=0$, apart from perturbative corrections.
To our knowledge, this model has not been considered in the literature.
Phase separation driven by a similar form of correlated hopping was found in \cite{Schmidt2006} for untruncated bosonic systems.
A pair checkerboard phase was predicted in \cite{Chen2011} for a system of bosons with three-body constraint for attractive contact interactions and repulsive nearest-neighbor interactions.
A stable pair superfluid phase 
at $\Delta/J \ll -1$
was suggested in \cite{Bonnes2011} for the case $\alpha=1$; this phase is stabilized by the three-body constraint, while the full Bose-Hubbard model displays a collapse for any $\Delta<0$.
Finally, the interplay between pair-hopping and a different form of correlated hopping has been considered in \cite{Eckholt2009},
in a cold atom implementation without three-body constraint.

The Gutzwiller results at unit density $n=1$ 
are reported in Fig.~\ref{fig:Gutzwiller_alfaDelta_fill1_main}. 
At this density, three thermodynamically stable phases are visible, a standard single-particle superfluid (SF), a pair superfluid (PSF) and the Mott insulator (MI).
To compactly illustrate the three phases, we plot $g^{(1)} - g_{\rm pair}^{(1)}$.

When $\Delta/J \leq -\frac{2z}{n}
    \left(
    \sqrt{n - n^2/2} + n \alpha
    \right)^2
 $, it is favorable to populate the state
$|2\rangle$, resulting in the formation of as many pairs as possible, i.e. $|\psi_2|^2=n/2$.
From a mean-field point of view, this is a pair superfluid phase, with $g^{(1)}=0$ and finite $g^{(1)}_{\rm pair}$.
However, we remark that the superfluid stiffness in this phase is determined by the effective pair-hopping constant obtained from second-order perturbation theory, which scales as $\sim \alpha^2 J^2/|\Delta|$. This means that the particles form  very heavy pairs and long-range coherence may be lost at very small temperatures. As a reminder of this fact, we label this regime with an extra asterisk, as PSF$^*$.

At large positive $\Delta/J \geq 
 z (1+\sqrt{2}\alpha)^2$,
instead, the ground state is a Mott insulator, with $\psi_1=1, \psi_0=\psi_2=0$; this can only occur at integer density $n=1$.
Both the PSF and MI are melted by the enhancement of (correlated) hopping, i.e. increasing $\alpha$, while exact solutions are obtained at zero $\alpha$.

\begin{figure}[t]
    \centering
\includegraphics[width=0.98\linewidth]{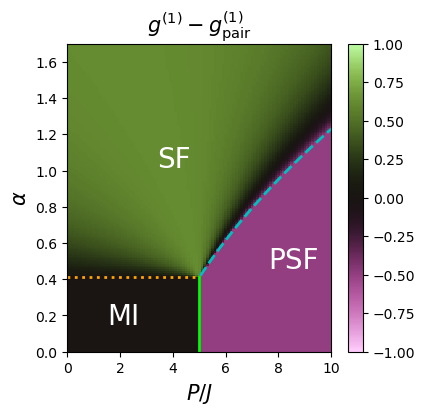}
    \caption{Difference between single-particle coherence $g^{(1)}$
    and pair coherence $g_{\rm pair}^{(1)}$
in the ground-state of $H$, as a function
of the hopping correlation $\alpha$ and of the pair-hopping strength $P/J$. Here, we take  $\Delta=2.5zJ$
and unit density $n=1$, and perform the calculations within the Gutzwiller approximation. We can distinguish a standard superfluid phase (SF, green region),  a pair superfluid (PSF, pink area) and a Mott insulator (MI, dark region).
The cyan dashed line separates the SF from the PSF, the orange dots the SF from the MI, and the solid light green line the MI from the PSF.
Analytical expressions for these transition lines are reported in the text.
}
\label{fig:Gutzwiller_alfaP_main}
\end{figure}

An instability towards cluster formation also occurs for large $\alpha/J$, for finite $\alpha$ and negative $\Delta$ 
and in a small region at very small $\alpha$. 
The instability is signaled by $\frac{d^2 e}{dn^2} <0$
and is represented by the red hatched area in the plots.
At large $\alpha$, this instability is driven by the correlated hopping term, since for $\alpha>1$ it is more favorable to hop in the presence of other particles. 
Reducing $\alpha$, and in particular around the standard bosonic hopping at $\alpha=1$, the collapse is driven by the Hubbard interaction term: for $\Delta/J$ sufficiently negative, the attractive interactions overcome the stabilizing effect of the kinetic energy and of the three-body  hard-core interaction.
Eventually, for very negative $\Delta/J$  the instability evolves into a stable gas of pairs. In Appendix F we use tensor network methods to estimate the instability range beyond the mean-field approximation (the $\alpha=1$  instability is actually suppressed at $n=1$~\cite{Cuzzuol2025}, but persists at lower density).

At small  $\alpha$, instead, it can be favorable (provided that $\Delta \lesssim 0$) to enhance the kinetic energy of  part of the particles by relegating the remaining particles in a high density region. This intriguing quantum phase separation of a Tonks-Girardeau gas and a dense solid of pairs is thoroughly described in Appendix F.

\subsubsection{Phase diagram in the  $(\alpha,P)$ plane}

We now consider the effect of pair-hopping processes $P \neq 0$
and their competition with the degree of hopping correlation 
$\alpha$.
Here, we consider the case of repulsive Hubbard interactions $\Delta=2.5 zJ$ and unit filling $n=1$.
The Gutzwiller results for $g^{(1)}-g^{(1)}_{\rm pair}$ are plotted in Fig.~\ref{fig:Gutzwiller_alfaP_main}.

The transition between the single-particle and pair superfluid is found along the line
\begin{equation}
P_{\rm crit}(\alpha)
=
\frac{2J}{n}
\left(
\sqrt{n-n^2/2} + n\alpha
\right)^2 + z^{-1}\Delta,    
\end{equation}
represented as the cyan dashed line.
Since here $P$ is the dominant scale, the PSF features light pairs and a large stiffness of order $P$;
this is in contrast with the heavy pairs previously found in   Fig.~\ref{fig:Gutzwiller_alfaDelta_fill1_main}, labeled as PSF$^*$.
Notice that PSF and PSF$^*$ are two regimes within the same phase.

The Mott transition occurs in the rectangle with height 
$\alpha_{\rm Mott} = \sqrt{\frac{\Delta}{2zJ}} - \frac{1}{\sqrt{2}}$, indicated by the horizontal orange dots,
and width $P_{\rm Mott} = 2z^{-1}\Delta$, given by the vertical green line.
The former line separates the MI from the standard superfluid via a second-order transition~\footnote{This refers to the mean-field result. Quantum fluctuations in 1D would instead lead to a BKT transition.}, while the latter transition is of the first order and divides the MI from the PSF.
Once again, the hopping correlation $\alpha$
plays a major role in the competition between different phases.
Notice that the repulsive interaction  greatly enhances the thermodynamic stability of the system; a cluster instability requires here a large value of $\alpha$ to occur.
A detailed study of this system at $\alpha=1$ and across the $(\Delta, P)$ plane has been provided in Ref. \cite{Malakar2023} using the Gutzwiller and cluster mean-field methods.

\subsubsection{Impact of the $\hat{W}$ term}

We will now investigate the impact of the  $\hat{W}$ term on the phase diagram of the qutrit simulator.
As explained in Section \ref{sec:general},
for inductive coupling we expect a generalized interaction term
$
 \hat{W}(\hat{\rho}_i,\hat{\rho}_j)
 = W(\hat{\rho}_i)W(\hat{\rho}_j)$ to be present,
 with
 $
 W(\hat{\rho}) = W_0  + \delta W_1 |1\rangle \langle 1 |
+ \delta W_2 |2\rangle \langle 2 |
$.
In the following, we take $W_0=0$,
since this just renormalizes the other terms in the Hamiltonian.
Indeed, in the representation
$
 \hat{W}(\hat{\rho})
 =
 W_0 + \tilde{W}_1 \hat{\rho} 
+
\tilde{W}_2 
(\hat{\rho}-1) \hat{\rho}$, it is more evident that the $W_0^2$ term is an overall energy shift, the $W_0 \tilde{W}_1$ term leads to a shift of chemical potential, and $W_0 \tilde{W}_2$
yields a Hubbard interaction term which renormalizes $\Delta$.
The most interesting terms are the $\tilde{W}_1^2$ term, which represents nearest-neighbor interactions,
and the $\tilde{W}_1 \tilde{W}_2$ and $\tilde{W}_2^2$ terms, which contain 3-body and 4-body operators, respectively. 

As we will see below, these (generalized) nearest-neighbor interactions can lead to translational symmetry breaking and to the formation of charge density waves.
To include these scenarios, we extend the Guzwiller ansatz to allow for different variational wavefunctions on different sites. In particular, we assume that the connectivity between qutrits is described by a bipartite lattice (e.g. a 1D chain or a 2D square lattice) with sites of type $C$ and $D$.
The density imbalance bewteen $C$ and $D$ thus provides an additional order parameter.

\begin{figure}[t]
    \centering
\includegraphics[width=0.98\linewidth]{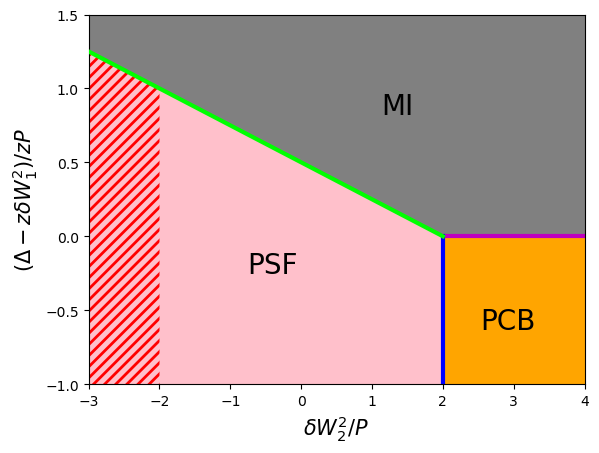}
    \caption{
    Phase diagram of the qutrit quantum simulator, in the regime where single-particle processes are not present, i.e. $J=0$, and at unit filling $n=1$.
    Within the Gutzwiller approximation,  three phases emerge:
    a Mott insulator (MI), a pair superfluid (PSF)
    and an incoherent pair checkerboard phase (PCB), which spontaneously break translational invariance. A phase separation instability can also occur (red hatched region).
}
\label{fig:Gutzwiller_W_impact}
\end{figure}

\begin{figure*}[t]
    \centering
\includegraphics[width=1.0\linewidth]{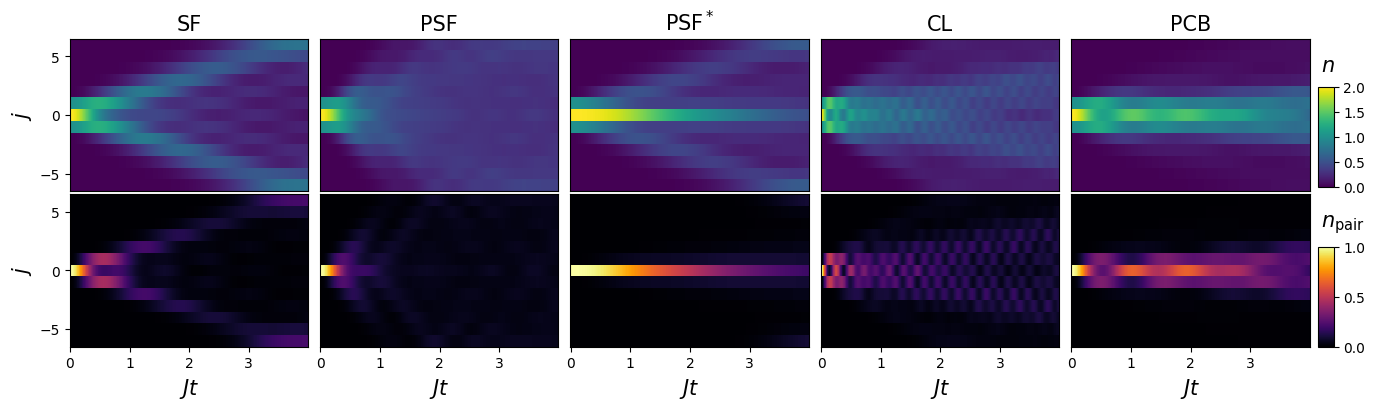}
    \caption{
    Space-time dynamics for an array of $L=13$ qutrits, starting from the initial state 
    $|\Psi(0)\rangle = \frac{1}{\sqrt{2}} b^\dagger_{-1} (b^\dagger_0)^2 b^\dagger_1 |0\rangle$.
    The top row illustrate the local density $n(j,t)$, while the bottom row report the density of pairs $n_{\rm pair}(j,t)$.
    Each column corresponds to different regimes (SF, PSF, PSF$^*$, CL, PCB) of the parameters in the Hamiltonian of Eq. (\ref{eq:the_H}), as detailed in the text.
}
\label{fig:dynamics}
\end{figure*}

Here, we focus on the simple case $J=0$ where single-particle hopping processes can be disregarded. This scenario can be achieved in the fluxonium qutrit regime and very similar Hamiltonians can be found in flat band systems, where some symmetry leads to destructive interferences (however, typically the hard-core three-body constraint is not present in these models).  
In this case, taking $J=0$ and  unit filling $n=1$, we find that three phases
can be realized for which analytical results exists at the Gutzwiller level.
The corresponding phase diagram is summarized in Fig.~\ref{fig:Gutzwiller_W_impact}.
The Mott insulator (MI in grey) and pair superfluid (PSF in pink) have already been introduced, since both phases are translationally invariant.
The PSF to MI transition occurs for
$\Delta -  z \delta W_1^2  = \frac{z}{2}P \left( 1 - \frac{\delta W_2^2}{2zP}
\right)$, corresponding to the light green line in the plot.
The third, yet to be discussed in this paper, phase is a pair checkerboard (PCB) phase, in orange in Fig.~\ref{fig:Gutzwiller_W_impact}.
This is an insulating state of pairs residing on one sublattice of the system.
Both single particle and pair coherences are zero, and the density imbalance is maximal.
The PCB to MI transition
occurs  for 
$\Delta=z \delta W_1^2$ (magenta line), which is independent of $\delta W_2$, while
the PSF to PCB transition occurs at
$P=
 \delta W_2^2/2$.
All these transitions are of the first order.
In Fig.~\ref{fig:Gutzwiller_W_impact} we sketch the phase diagram
as 
function of $\delta W_2^2/P$
and $(\Delta -  z \delta W_1^2)/zP$.
We clearly see that  $\delta W_2$ has a strong effect on the pair superfluid phase.
There is also a red hatched region for 
$2P+\delta W_2^2 < 0$ where the PSF becomes unstable because of the attractive nearest-neighbor interactions between the pairs.

\subsubsection{The $J\neq0, P<0$ case.}

In the previous paragraphs, we considered the situation where both $J$ and $P$ are non-negative, $J,P \geq 0$.
As explained in Appendix A, in a bipartite lattice or in a chain with open boundary conditions,
the physics in the $J<0,P>0$ case is essentially equivalent.

The scenario with negative $P$ and real non-zero $J$
is instead qualitatively different.
As investigated in Refs.~\cite{jurgensen2015twisted,Luhmann}
in the absence of the 3-body constraint,
the two kinetic terms lead to a frustration for the phase of the superfluid over different sites.
At the mean-field level, one would use the ansatz
$\langle \hat{b} \rangle_C = \sqrt{n} e^{i\theta/2}, \langle \hat{b} \rangle_D = \sqrt{n} e^{-i\theta/2}$, leading to a kinetic energy contribution $-J n \cos\theta - P\frac{n^2}{2} \cos2\theta$.
The minimization of the  single particle hopping term 
requires $\theta=0$ for $J>0$ or $\theta=\pi$ for $J<0$, while $\theta=\pi/2$ is required for pair hopping. 
Notice that when $|P/J| \gg 1$, the ground state consists of a pair-superfluid with $k=\pi$ momentum.
The Gutzwiller study of the full bosonic system revealed 
a time-reversal spontaneous symmetry breaking to
a twisted superfluid phase with 
$0<\theta<\pi/2$ for $J>0$ 
(or $\pi/2<\theta<\pi$ for $J<0$)
at intermediate values of $P/J$~\cite{jurgensen2015twisted}.

The application of the  Gutzwiller ansatz to the qutrit system instead yields a sharp transition from a standard superfluid to a $k=\pi$ pair superfluid, but the 3-body constraint suppresses the twisted superfluid phase (not shown).
Some preliminary exact diagonalization results are reported in Appendix F and reveal complex physics with strong fluctuations 
at an intermediate value of $P/J$.

A full investigation of qutrits with $J\neq0, P<0$ will require further studies, including the exploration of alternative lattice geometries, such as dimerized lattices where interactions can induce spontaneous fluxes and chiral currents~\cite{Goldman2023}.

\subsection{Quantum dynamics}

In the previous pages, we described the rich phase diagram of the qutrit Hamiltonian of Eq.~(\ref{eq:the_H})
at the ground state level.
However, superconducting circuits are open quantum systems, subject to decoherence and losses, so it is a priori not clear whether it is possible to engineer a bath capable to stabilize the different many-body phases, as was achieved for a MI phase in Ref.~\cite{ma2019dissipatively}.
Here, we will provide simple examples of dynamical experiments which can be straightforwardly performed in the lab, on timescales shorter than the microsecond coherence times estimated in Eqs. (\ref{eq:Tdepola},\ref{eq:Tdeph}).
The system is initialized in a tensor product of eigenstates of the local number operator, and the dynamics is probed in time by measuring this same operator.
Importantly, these operations have been realized experimentally in several landmark studies probing excitation dynamics in transmon arrays~\cite{roushan2017spectroscopic,yan2019strongly,karamlou2022quantum,braumuller2022probing,Chiaro2022}, but have yet to be realized in fluxonium arrays.
Following this protocol leads to qualitatively different dynamical behaviors in the different parameter regimes of $\hat{H}$, as we will now see. 

Our results are summarized in Fig.~\ref{fig:dynamics}.
We study the Hamiltonian time-dynamics for a open boundary system of $L=13$ qutrits initialized in a state $|\psi(0)\rangle$ with 4 photons at the center of the array. More precisely,
we take the product state
$|\Psi(0)\rangle = \frac{1}{\sqrt{2}} b^\dagger_{-1} (b^\dagger_0)^2 b^\dagger_1 |0\rangle$.
We numerically obtained the evolved state
$|\Psi(t)\rangle) = e^{-iHt} |\Psi(0)\rangle $
and evaluated the local density 
$n(j,t) = 
\langle \Psi(t)|
\hat{n}_j
|\Psi(t)\rangle$
and the local doublon density 
$n_{\rm pair}(j,t) = 
\langle \Psi(t)|
\hat{n}_j (\hat{n}_j-1)
|\Psi(t)\rangle$.
In Fig.~\ref{fig:dynamics}, we plot $n(j,t)$ and $n_{\rm pair}(j,t)$ as a function of space and time on the upper and lower row, respectively.
In each of the five columns, we consider a different regime of parameters of $\hat{H}$, indicated by the top label.

Starting from the left-hand side, we consider the single particle superfluid regime (SF), where $J$ is finite, $\alpha=1$ and all other parameters are zero, $P=\Delta=W_r=0$.
As expected, the particle disperse, resulting in a V-shaped density dynamics. The density of pairs strongly correlates with the total density.

Next, we considered the pair superfluid regime (PSF) with 
$P=3J$. The first observation is that the spread of the density occurs on a shorter timescale, driven by the motion of pairs with a large pair-hopping constant. The pair-hopping term is a many-body term and it turns out that the hopping of pairs is characterized by a double V-shaped pattern, visible in $n_d(j,t)$. Because of the presence of both single particle and pair hopping, $n(j,t)$ and $n_{\rm pair}(j,t)$ correlate only partially, $n(j,t)$ being significantly more blurred.

In the central column we find the regime of small $\alpha = 0.4$ and attractive Hubbard interaction $\Delta = -2 J$. This corresponds to the pair superfluid with small stiffness identified above, and we indicate it in the plot as PSF$^*$. 
While single particles can diffuse freely, the central pair is energetically bound by $\Delta$ and is also stabilized by the small $\alpha$, which unfavors dissociation. Since $P=0$, this pair disperses with a slow rate $|\alpha^2 J^2 / \Delta|$.

Then, we have the cluster regime (CL) with $\alpha=2$ and $P=\Delta=W_r=0$.
For these large values of $\alpha$, the system displays collapse instabilities towards a high-density phase with large kinetic energy, enhanced by the correlated hopping. This instability is more evident at small densities, see the red hatched region in Fig~\ref{fig:Gutzwiller_alfaDelta_App}. 
At the dynamical level, the tendency of the system towards collapse manifests itself in a certain resilience against dispersing and in high-frequency oscillations.

Finally, the leftmost column illustrates the pair checkerboard regime (PCB)
with 
$P=J, w_1=0.7, w_2=4$.
Notice that $|\Psi(0)\rangle$ does not contain the typical correlations of the ground-state of the system at half-filling, on the contrary this is  a very high energy state. 
While a fraction of the single particle amplitude disperses with rate $J$, the dynamics of pairs  stays confined at the center of the systems and displays oscillations.

Our results demonstrate that different terms and regimes of the Hamiltonian give rise to distinctive dynamical signatures, which can be probed in experiments.
A full treatment of the open quantum system, the stabilization of ground states, and the evolution of more complex states will be the topic of future research.

\section{Conclusions and outlook}
\label{sec:outlook}

In summary, we have theoretically demonstrated that, with current technology,  qutrit quantum simulators can be realized in arrays of fluxonium circuits.
At the individual qutrit level, by tuning the magnetic flux bias in a single device,
four different regimes can be achieved, classified by the  plasmon-like versus fluxon-like nature of the qutrit transitions.
At the array level,
we have provided a general mapping of the circuit to a bosonic Hamiltonian, containing correlated and pair hopping processes, as well as local and non-local interactions.
This Hamiltonian possesses a rich ground state phase diagram, comprising Mott insulating, crystalline, clustered,
superfluid and pair superfluid phases.
We have also proposed simple dynamical experiments, displaying  distinctive patterns in the different regimes.
We have estimated the impact of losses and decoherence to be comparable to that of transmon devices; we have also considered off-resonant photon exchanges that can in principle undermine our treatment of qutrits and have computed perturbative corrections;
our results indicate that adopting couplings in the window of 5-50 MHz allows one to safely operate the qutrit simulator.

The many-body physics explored in this work represents a first glimpse of the possibilities opened by qutrit-based quantum simulators. Their combination of strong interactions, multi-level structure, and tunable couplings
paves the way to a number of  future developments.

A first and natural extension concerns the application of the platform to topological phases and quantum Hall physics. 
In particular, 
 the Floquet engineering techniques demonstrated in Refs.
 \cite{Roushan2016chiral,Wang2024fqhe,Rosen2024}, if used in the context of qutrits and qudits,
 would provide a strategy to separately control the amplitude and Peierls phases of each hopping channel.
 This can be used to 
  generate synthetic electric and magnetic fields or implement density-dependent gauge potentials. In principle, this is a 
  powerful toolbox, which
  would enable the realization of non-Abelian topological order~\cite{Wen2007}, lattice gauge fields~\cite{Banuls2020} and anyon-Hubbard-type models~\cite{Keilmann2011}. 
  A microscopic analysis of this engineering strategy will be soon carried out.
  Few-qutrit experiments, in the spirit of the three-qubit characterization of Roushan \emph{et al.}~\cite{Roushan2016chiral}, would provide valuable insight and proof-of-concepts in the near future.

In general, the role of the pair-hopping term $P$ on the stability of (non-Abelian) quantum Hall phases, such as the Pfaffian state, also deserves attention. Indeed, understanding how the various terms in the Hamiltonian $\alpha$, $P$, and $W$ -- neglected in earlier proposals~\cite{Hafezi2014Engineering}-- affect its robustness is a natural application of the tunability offered by qutrit devices. 

In principle, correlated pair tunneling can give rise to twisted superfluid phases in Hubbard-like systems, characterized by a complex order parameter that encodes a non‑trivial relative phase between neighboring sites~\cite{jurgensen2015twisted,Luhmann}. It remains to be investigated under which conditions (e.g. in which lattices) such a phase can be stabilized within qutrit simulators. Moreover, realizing twisted superfluidity on a dimerized lattice, where pair hopping occurs within each dimer, can produce striking emergent effects, including spontaneous (interaction‑induced) fluxes and chiral currents~\cite{Goldman2023}. The high degree of tunability offered by qutrit circuit‑QED platforms positions them as an ideal setting to uncover these phenomena linked to the spontaneous breaking of time‑reversal symmetry.

Geometrically frustrated lattices provide further opportunities. The $\pi$-flux diamond chain, which was implemented with superconducting circuits in Ref.~\cite{Martinez2023}, features  a set of flat single particle bands hosting, in the weakly interacting limit, interesting phenomena like 
phases with chiral order~\cite{DiLiberto2023},
emergent gauge constraints, self-pinning~\cite{Burgher2025}, and disorder-free localization~\cite{Brenes2018}. This physics is  strongly modified in qubit arrays, due to the Hilbert space truncation; it would be interesting to investigate the scenario in qutrit simulators,
by exploiting the tunable on-site interaction.

Non-equilibrium physics is another frontier. For instance, the PSF--SF transition is believed to be first order~\cite{Zhang2013PSF_QMC}, and, at finite temperature, the predicted BKT transition would be driven by fractional $\pm 1/2$ vortices~\cite{Ng2012}, reflecting the reduced $U(1)/\mathbb{Z}_2$ symmetry breaking.
Yet, the dynamical signatures of this transition remain largely unexplored and would be especially interesting to observe in a strongly interacting quantum regime.

The platform also naturally connects to quantum magnetism. By tuning the qutrit parameters, the simulator can in principle realize  spin-1 SU(2) and SU(3)-symmetric  Heisenberg models. Given the ability to tune the sign and structure of inter-qutrit couplings, this architecture could become a useful tool for exploring quantum spin liquids~\cite{Wen2007}.

Finally, practical considerations such as decoherence and losses merit systematic study. For instance, when $J=0$ and $P\neq 0$, single-particle losses couple different gauge sectors and impose strong dynamical constraints. Developing efficient preparation protocols -- including shortcuts to adiabaticity, dissipative schemes, and optimal-control approaches -- will also be essential for reliably accessing exotic ground states such as the Pfaffian.

\section{Acknowledgments}

We are grateful to Denis Basko and Julien Renard for useful disussions.
I.A. and N.G. acknowledge financial support by the ERC grant LATIS, the EOS project CHEQS, the FRS-FNRS (Belgium)
and  the Fondation ULB. 
Q.F. acknowledges financial support by the French grant ANR-22-PETQ-0003 under the ``France
2030 plan'' as well as by the European Union by the EU Flagship on Quantum Technology HORIZON-CL4-2022-QUANTUM-01-SGA project
101113946 OpenSuperQPlus100.
%%%%%%%%%%%%%%%

\vspace{2cm}

\appendix

\begin{figure}[b]
    \centering
\includegraphics[width=0.95\linewidth]{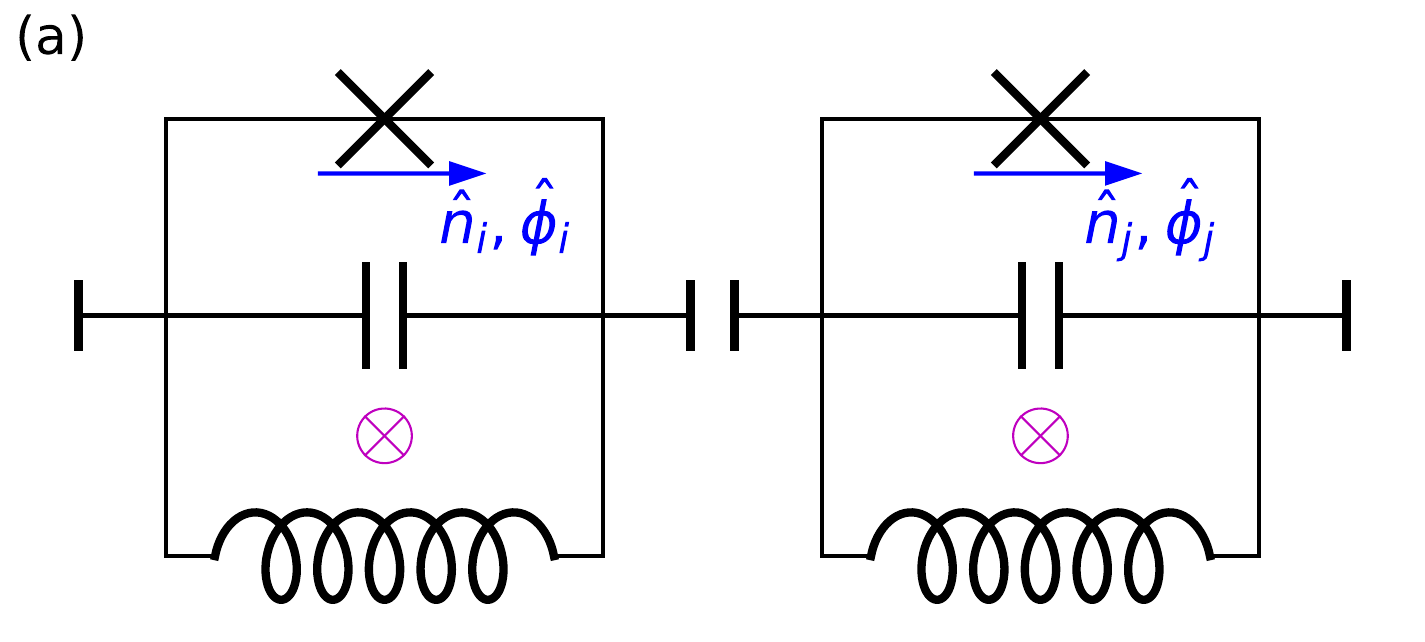}
\includegraphics[width=0.95\linewidth]{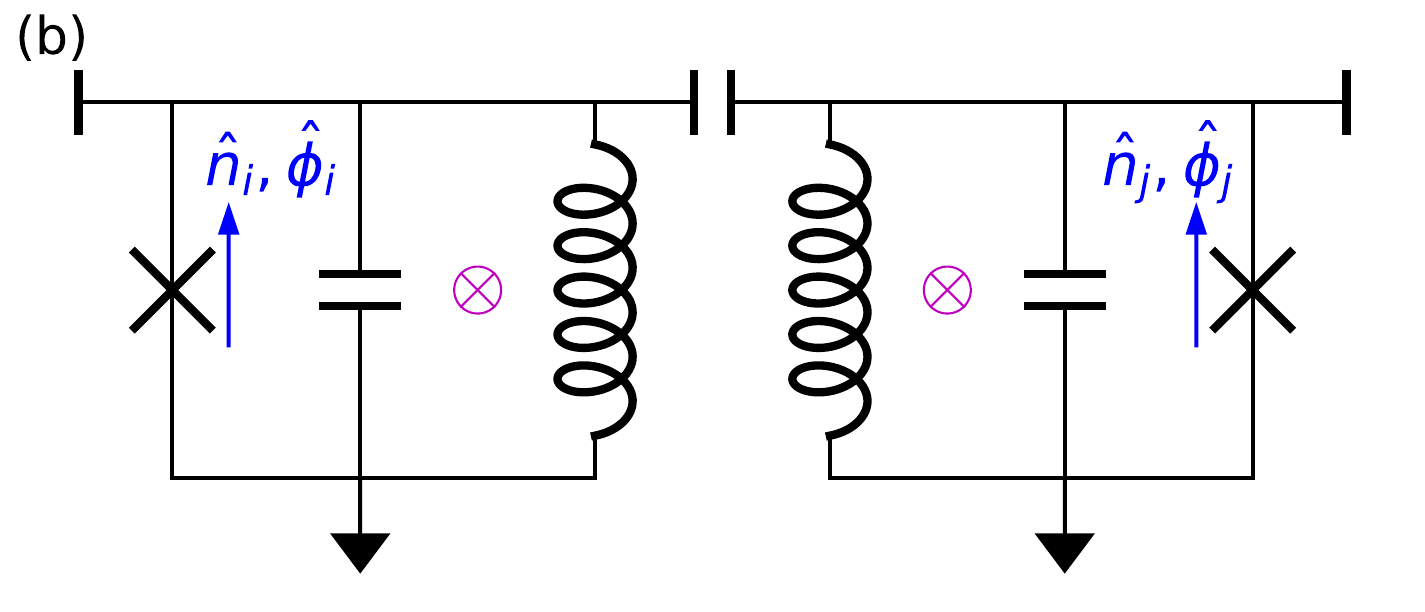}
    \caption{
    Two capacitative coupling schemes between fluxonium atoms leading to effectively negative coupling.
    In particular, the circuit in (b) maps to 
    Eq. (\ref{eq:Hfig8}) via the transformation 
    $\hat{\phi}_j \to (-1)^j \ \hat{\phi}_j,\hat{n}_j \to (-1)^j \ \hat{n}_j$.}
    \label{fig:negative_gC}
\end{figure}

\section*{Appendix A: Capacitative coupling schemes}

In this Appendix, we aim to provide some considerations regarding realistic coupling schemes between fluxonium atoms, with a special focus on the sign of the interaction term occurring in the Hamiltonian. We will focus on the case of capacitative coupling, but we expect that similar conclusions can be drawn for inductive coupling.

Let's consider first the infinite linear chain of capacitatively coupled circuits 
sketched in Fig.~\ref{fig:sketch} of the main text.
Before quantization, 
this is
described by the Lagrangian~\cite{2021Rasmussen}
\begin{equation}
\mathcal{L}[\vec{\phi},\dot{\vec{\phi}}]
=
\frac{1}{2} \sum_{ij} 
C_{ij} \dot{\phi}_i \dot{\phi}_j - V(\vec{\phi}), 
\end{equation}
with $C_{ij}=C_q\delta_{i,j} - 2C_c\delta_{i,j+1}
 - 2C_c\delta_{i,j-1}$.
Here, $C_q$ is denotes the capacitance of the capacitor connecting the two islands within each fluxonium circuit, while $C_c$ refers to the capacitors connecting different artificial atoms. 

The corresponding Hamiltonian in terms of the conjugate momenta $n_i=\sum_jC_{ij}\dot{\phi}_j$ reads
\begin{equation}
   H[\vec{\phi},\vec{n}]
=
\frac{1}{2} \sum_{ij} 
C_{ij}^{-1} n_i n_j + V(\vec{\phi}) 
\end{equation}
and can be canonically quantized.
The inverse capacitance matrix can be easily obtained by Fourier transformation and series expansion to obtain
\begin{equation}
     C^{-1}_{j,j+r} =  \frac{1}{C_q+2C_c} \sum_{m=0}^\infty 
\binom{r+2m}{m}
\left( 
\frac{C_c}{C_q+2C_c}
\right)^{r+2m}.
\end{equation}
In the limit $C_c \ll C_q$, 
the inverse capacitance matrix behaves at leading order as
$C_{j,j+r}^{-1}
\simeq \frac{1}{C_q+2C_c} [C_c/(C_q+2C_c)]^r
$,
with $r \geq 0$ labeling the $r$-th nearest-neighbor coupling. Since in this paper we are interested in a parameter regime where $g_C/E_C\simeq C_c / C_q<1/10$,
we neglect the terms with $r\geq 2$. While we expect that these terms would provide very small quantitative corrections to the results discussed in this paper, in the case of geometrically frustrated systems, where there is a close competition between different states, these corrections may play a non-negligible role in suppressing or favoring a particular phase.

\begin{figure*}
    \centering
    \includegraphics[width=0.48\linewidth]{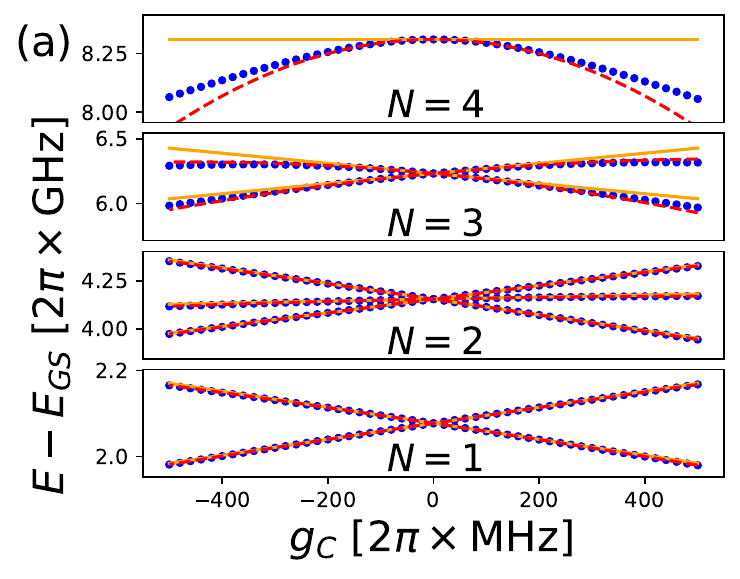}
    \includegraphics[width=0.48\linewidth]{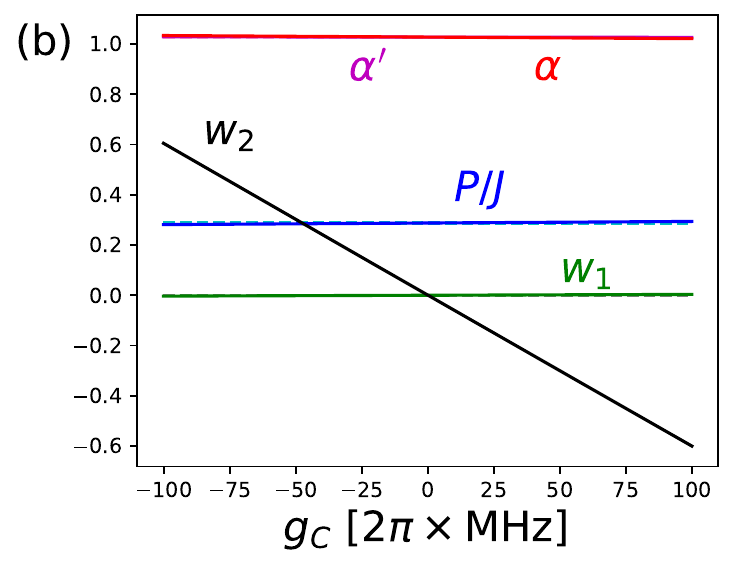}
    \includegraphics[width=0.48\linewidth]{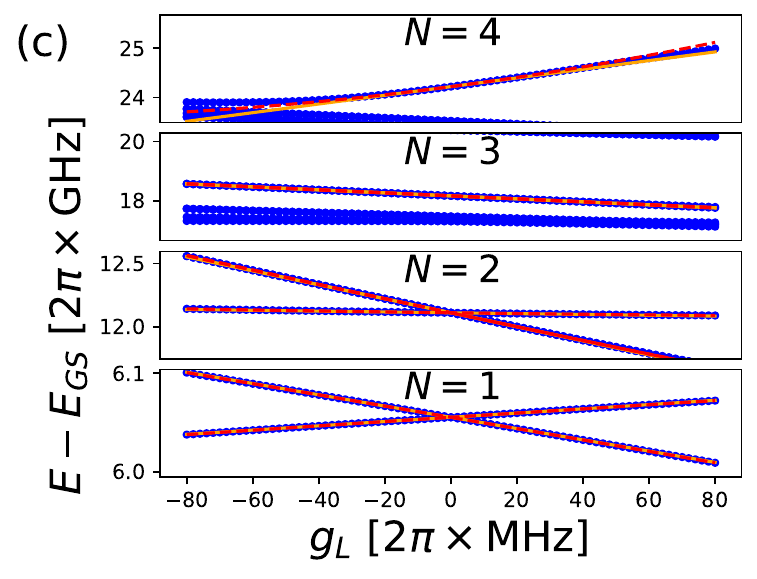}
    \includegraphics[width=0.48\linewidth]{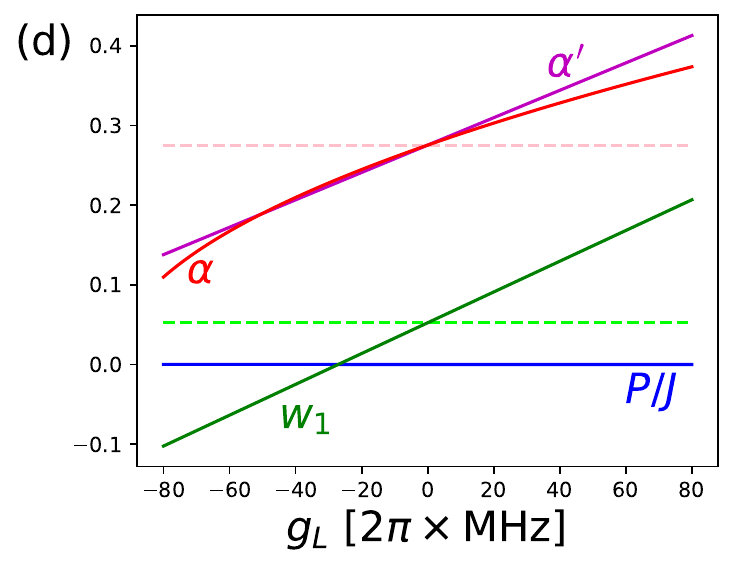}
    \caption{
    Perturbative corrections to the spectrum and effective interactions between two qutrits, arising from non-resonant processes involving high-energy levels. 
    (a) We diagonalize the full spectrum of two coupled fluxonium circuits as a function of the strength of the capacitative coupling $g_C$, and we plot the eigenvalues as blue points. The predictions obtained through the $9 \times 9$ two-qutrit Hamiltonian of Eq. (\ref{eq:the_H}) are displayed by the orange lines. The red dashed lines include the non-resonant corrections $\delta H$ at the second-order of perturbation theory, see Eq. (\ref{eq:dH}). These corrections can be rephrased in terms of the renormalization of the qutrit-qutrit interaction parameters $\alpha, P/J, w_1, w_2$, as illustrated in panel (b), where the dashed horizontal lines refer to the RWA expectations. While panels (a,b) refer to the most robust \plasmonium~regime and capacitative coupling, panels (c,d) deal with the most sensitive \plasmofluxonium~regime and inductive coupling, for which the two-qutrit truncation of the Hilbert space works well for $g_L$ only up to a few tens of MHz.
    In panel (d), $w_2$ is of the order of 50 and it is out of the plot scale.
    }
    \label{fig:dispersive_corrections}
\end{figure*}

\begin{figure*}
    \centering
\includegraphics[width=0.98\linewidth]{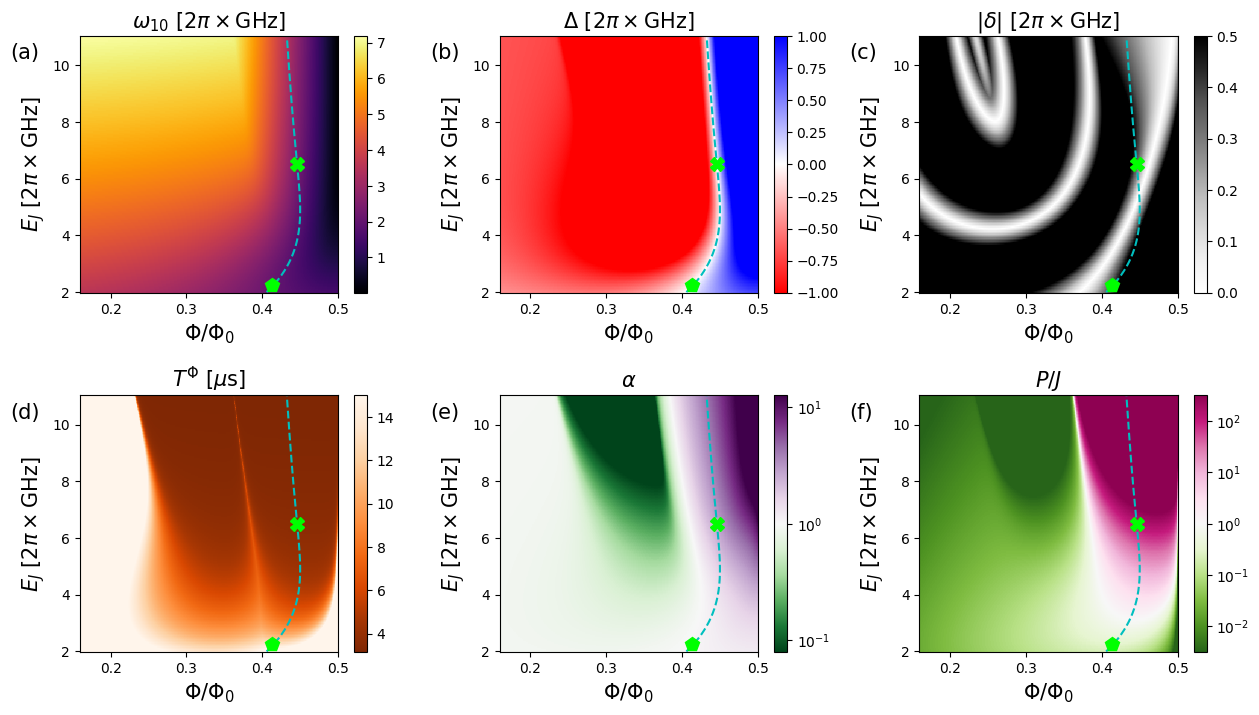}
    \caption{
    Qutrit parameters as a function of $\Phi$ and $E_J$, when the 
    qutrit is made from the three lowest levels of $H_{\rm at}$, for a fluxonium circuit. 
    The cyan dashed line indicates the qutrit condition $\Delta=0$ being satisfied.
    The  green pentagon and cross refer to the  parameters of the  $\Pi\Pi$ and $\Phi\Phi$ qutrits reported in Fig.~\ref{fig:inductance_qutrits}.(a) and (b).
    We fixed $E_C=0.60$~GHz and $E_L=1.5$~GHz.
    }
    \label{fig:6panel_123}
\end{figure*}

We have thus recovered the  capacitative coupling Hamiltonian $\hat{H}_C^{ij} = g_C \hat{n}_i \hat{n}_j$
introduced in the main text, with $g_C>0$ for this configuration.
In Fig.~\ref{fig:negative_gC},
we analyze two circuit configurations for which $g_C<0$ instead.

In panel (a), the right island of the  fluxonium $i$ on the left is capacitatively coupled to the left island of the fluxonium $j$ on the right.
The idea is that, while the coupling between the charge fluctuations in these two islands is positive (an excess of charge in both islands leads to an increase of energy), the convention for the operator $\hat{n}_j$ refers to the charge excess of the right island of atom $j$, whose fluctuations are opposite in sign to the ones of the left island, since charge is conserved.
The coupling constant in $\hat{H}_C^{ij} = g_C \hat{n}_i \hat{n}_j$
is thus negative, $g_C<0$.
This strategy has been beautifully exploited in Ref.~\cite{Martinez2023} to realize a $\pi$-flux square plaquette with transmons, where one side featured $J<0$  and three sides 
$J>0$.

It is worth considering also the configurations displayed in Fig.~\ref{fig:negative_gC}.(b), relying on grounded circuits (as a technical remark, grounding can provide some practical advantage). To fully appreciate this strategy, we refer to the next subsection, where the importance of the sign of $P$ (rather than the sign of $J$) and the role of bipartite lattices are stressed.
Here, for simplicity we consider an open chain of fluxonium atoms, where for even $j$ the circuit structure is specular to that of odd $j$.
This is physically equivalent to having identical circuit but a staggered magnetic flux (a situation more difficult to realize in practice).
The Hamiltonian reads
\begin{equation}
    \hat{H}
    =
    \sum_j \hat{H}_{\rm at}[\hat{\phi}_j,\hat{n}_j,(-1)^j \Phi]
    + 
    g_C \hat{n}_j \hat{n}_{j+1},
\end{equation}
with $g_C>0$.
Applying the transformation
$\hat{\phi}_j \to (-1)^j \ \hat{\phi}_j,\hat{n}_j \to (-1)^j \ \hat{n}_j$
leads to 
\begin{equation}
    \hat{H}
    =
    \sum_j \hat{H}_{\rm at}[\hat{\phi}_j,\hat{n}_j, \Phi]
    - 
    g_C \hat{n}_j \hat{n}_{j+1}.
    \label{eq:Hfig8}
\end{equation}
In getting the last equation, we relied on the fact that in fluxonium the parity breaking term is provided by the flux.
This is really the crucial point, since the $P = - g_C |n_{02}|^2$  is non-zero only in the presence of the parity breaking field. Similar considerations apply, for inductive coupling, to the $\hat{W}$ term.
In conclusion, the physics of the circuit depicted in Fig.~\ref{fig:negative_gC}.(b) is  intrinsically different from the physics of the circuit of Fig.~\ref{fig:sketch},
described by 
\begin{equation}
    \hat{H}
    =
    \sum_j \hat{H}_{\rm at}[\hat{\phi}_j,\hat{n}_j, \Phi]
    + 
    g_C \hat{n}_j \hat{n}_{j+1}.
    \label{eq:Hfig1}
\end{equation}
Notice that in an open chain of transmons, whose Hamiltonian does not contain the magnetic flux, the Hamiltonians in Eqs. (\ref{eq:Hfig1}) and (\ref{eq:Hfig8}) describe instead the same physical system, with two different definitions of the phase and charge variables.

\subsubsection*{Considerations on the sign of $J$ and $P$}

We have thus demonstrated the possibility of experimentally realizing both 
$g_C>0$
and
$g_C<0$, at least for bipartite lattices.
We recall that, with the gauge choices introduced in Sec. \ref{sec:general},
the sign of $J,P$ is the opposite of the sign of $g_C,g_L$, while (for inductive coupling) the nearest-neighbor interactions are repulsive or attractive  for  positive or negative $g_L$, respectively.
Here, we would like to further consider the consequences of the sign of $J$ and $P$ at the level of the bosonic Hamiltonian of Eq. (\ref{eq:the_H}).
In a bipartite lattice with sublattices $C,D$ (this includes open one-dimensional chains), the
sign of $J$ is not essential.
Indeed, we can always perform the unitary transformation $\hat{b}_j \to s_j \hat{b}_j$,
where $s_j=1$ if $j\in C$ and $s_j=-1$ for $j\in D$.
This transformation leaves all density correlators invariant and introduces some minus signs in the single-particle coherence function, without changing its absolute value. The pair coherences are also unaffected.
At the level of the Hamiltonian, the only change is the sign of $J$, which is flipped.
In other words, the ground states of the two Hamiltonians containing $J$ or $-J$ have very similar properties. For example, if the ground state of the first is a Mott state, the same holds for  the second one; if  the first system is superfluid, the ground state of the second one is a boosted superfluid with momentum $k=\pi$.
Clearly, this transformation is not useful on closed chains with $L$ odd or threaded by a finite flux.

If $J=0$, similar considerations hold for the sign of $P$, as obtained from the transformation  with $s_j=i$ for $j\in D$.
However, if both $J$ and $P$ are nonzero and if $J$ is taken real, the sign of $P$ leads to
very different physics in the $P>0$ or $P<0$ cases, as anticipated in Sec. \ref{sec_many-body}.

\section*{Appendix B: perturbative corrections}

When a qutrit is coupled to other qutrits, cavity resonators and external control lines, in principle one should diagonalize the whole circuit, containing many degrees of freedom. In practice, if the coupling strength to the other elements is weak with respect to the typical qutrit energy scale $\omega_{10}$, the analysis leading to $\hat{H}$ of Eq. (\ref{eq:the_H}), based on the RWA, is  a good starting point.
Perturbation theory can be used to calculate corrections to Eq. (\ref{eq:the_H}), which in many experiments are known to lead to measurable
dispersive
shifts of a few MHz~\cite{zhu2013}.
These corrections originate from non-resonant, anti-rotating wave, processes, where intermediate states violate the conservation of the photon number.

We evaluated such corrections  
for two coupled qutrits $i$ and $j$ using the Schrieffer-Wolf method at the second-order perturbation theory level.
The qutrit subspace is defined by the indices $a,b,c,d \in \{  0,1,2 \}$ and the RWA $a+c=b+d$
is assumed.
The Hamiltonian of Eq. (\ref{eq:the_H}) 
can be expressed as a $9\times9$ matrix $H_{ac,bd}$, whose terms are to be renormalized by the perturbative corrections
\begin{equation}
    \delta H_{ac,bd} = 
g_C^2 \sum_{\bar{r} \bar{s}}
\frac{
 \langle a | \hat{n}_i | \bar{r} \rangle \langle c | \hat{n}_j | \bar{s} \rangle
\langle \bar{r} | \hat{n}_i | b \rangle \langle \bar{s} | \hat{n}_j | d \rangle
}{\omega_a+\omega_c-\omega_{\bar{r}}-\omega_{\bar{s}}}.
\label{eq:dH}
\end{equation}
Here $\bar{r}, \bar{s}$ span the whole artificial atom Hilbert space, with the only constraint that the denominator be nonzero (corresponding to non-resonant photon exchanges).
For brevity, in Eq. (\ref{eq:dH}) we only considered capacitative coupling, the extension to include inductive coupling being straightforward.
Also, in writing the coupling Hamiltonian as 
$\hat{H}_C^{ij} = g_C \hat{n}_i \hat{n}_j$,
we implicitly absorbed the self-capacitance corrections due to the coupling capacitor into the definition of $\hat{H}_{\rm at}$.

Typical results are reported in Fig.~\ref{fig:dispersive_corrections}.
In panel (a), we plot the spectrum of two capacitatively coupled qutrits in the plasmon-plasmon regime, as a function of $g_C$. 
The blue points have been obtained through the exact diagonalization of the whole circuit, fully keeping into account the phase and charge operators of both qutrits. 
The orange lines, instead, are the eigenvalues of the two qutrit Hamiltonian of Eq. (\ref{eq:the_H}), while the red dashed lines also keep into account the perturbative corrections of Eq. (\ref{eq:dH}).
To optimize the plot, we zoom in close to the energy of the uncoupled qutrits in each photon number sub-manifold $N=1,2,3,4$.

In panel (b), the qutrit parameters $J, \alpha, P, W_r$ appearing in Eq. (\ref{eq:the_H}) are calculated both following the zeroth order recipe of Section \ref{sec:general} (dashed lines) and including the perturbative corrections (solid lines).
More precisely, after defining the $9 \times 9$  Hamiltonian matrix $H' = H + \delta H$, we extract the single particle hopping as $J=H'_{01,10}$, the pair hopping as $P=H'_{02,20}$
and $\delta W^2_r = H'_{rr,rr} - 2H'_{0r,0r} +H'_{00,00}$.
To remove the trivial proportionality on $g_C,g_L$, we plot $P/J$ and $w_r=-\delta W^2_r/(J+P)$. Since with our definitions ${\rm sign}(J)={\rm sign}(P)=-{\rm sign}(g)$, the sign of the physical interactions 
is given by ${\rm sign}(\delta W^2_r)
= {\rm sign}(w_r) \cdot {\rm sign}(g)$.
Concerning the hopping correlation, we compare the two definitions $\alpha = H'_{11,20}/ (\sqrt{2}J)$ and 
 $\alpha' = \frac{1}{2}\sqrt{H'_{21,12}/J}$; in the plot, $\alpha$ and $\alpha'$ are represented by the magenta and red lines respectively, and they remain quite close to each other.

 \begin{figure*}
    \centering
\includegraphics[width=0.98\linewidth]{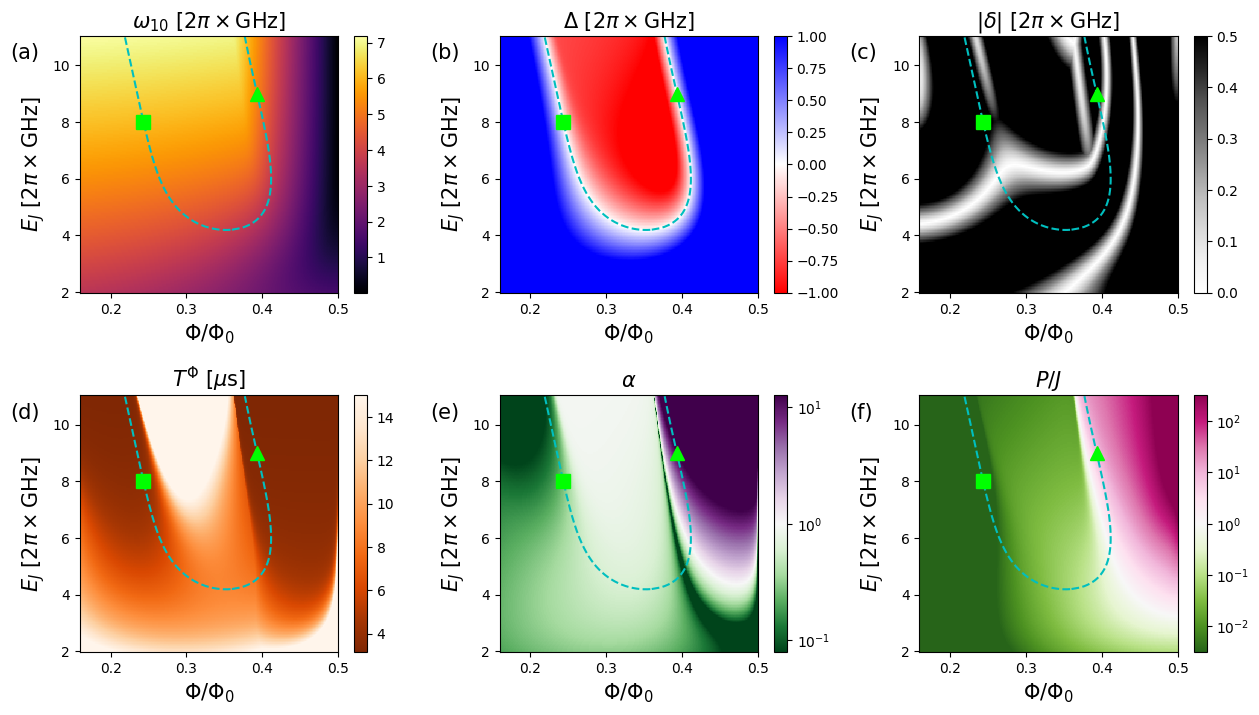}
    \caption{ Qutrit parameters as a function of $\Phi$ and $E_J$, when the 
    qutrit is made from the first, second and fourth levels of a fluxonium circuit, i.e. $|0\rangle=|\bar{0}\rangle,|1\rangle=|\bar{1}\rangle,|2\rangle=|\bar{3}\rangle$.}
    \label{fig:6panel_124}
\end{figure*}

The operative regime of the  \plasmonium~makes it pretty resilient to the inter-qutrit coupling, and dispersive corrections are negligible up to tens of MHz of $g_C$, with second-order perturbation theory working excellently up to $g_C \sim 200$~MHz.
At these large values of $g_C$, the emergent non-local interactions between pairs, quantified by $w_2$, are attractive and sizable, and may eventually lead to collapse instabilities.
Also, notice that the spectrum reported in panel (a) is slightly asymmetric in $g_C$, reflecting the weak asymmetry of the $V(\phi)$ in the $\Pi\Pi$ regime.

In panels (c) and (d) 
we report the analogous data of (a) and (b) for the most sensitive case of
the inductive coupling of two {\plasmofluxonium}s.
Here, the validity of our framework is limited to $g_L$  of a few tens of MHz.
In particular, the presence of the spurious level $\bar{2}$ leads to some   avoided crossing between states within and outside of the qutrit subspace ($N=4$ subplot, at negative $g_L$).
Finally, in this regime, the asymmetry in $g_L$ is sizable. 

In conclusion, while these perturbative corrections complicate the theoretical analysis of the device, the physics is qualitatively captured by the RWA Hamiltonian of Eq.~(\ref{eq:the_H}), provided couplings of few tens of MHz are used. Keeping into account the coherence times of a few $\mu$s, an operational  window of 5-50 MHz for the coupling strengths seems optimal.

\section*{Appendix C:  Exploration of qutrit parameters in fluxonium circuits}

The goal of this Appendix is to provide a more comprehensive study of the behavior of the qutrit parameters when 
$\Phi$ and $E_J$ are varied. For the sake of simplicity and readability, we have instead fixed $E_C=2\pi\times0.60$~GHz and $E_L=2\pi\times 1.5$~GHz.
This choice is also motivated by experimental realizations, where $\Phi$ and $E_J$ are more easily adjusted than 
$E_C$ and $E_L$.

In Fig.~\ref{fig:6panel_123},
we consider a qutrit made from the three lowest levels of $H_{\rm at}$, that is 
$|0\rangle=|\bar{0}\rangle,|1\rangle=|\bar{1}\rangle,|2\rangle=|\bar{2}\rangle$
in our notation.
In Fig.~\ref{fig:6panel_123}.(a) we report the frequency of the first transition,
while in panel (b) we plot the detuning $\Delta=\omega_{21}-\omega_{10}$ of level 2,
which enters as a Hubbard term in the Hamiltonian of Eq. (\ref{eq:the_H}).
The cyan dashed  curve corresponds to the resonant  condition $\Delta=0$.
The absolute value of $\delta$, the detuning
to the closest extra level, is illustrated in panel (c).
Panel (d) report the coherence time against flux noise decoherence, as calculated in Eq. (\ref{eq:Tdeph}), while panels (e) and (f) illustrate $\alpha$ and $P/J$ for capacitative coupling.
The light green pentagon and cross present in all the panels refer to the  parameters of the  $\Pi\Pi$ and $\Phi\Phi$ qutrits reported in Fig.~\ref{fig:inductance_qutrits}.(a) and (b), respectively.

Figure \ref{fig:6panel_124} reports exactly the same quantities but for a qutrit consisting of  the first, second and fourth levels of  $H_{\rm at}$, i.e.
$|0\rangle=|\bar{0}\rangle,|1\rangle=|\bar{1}\rangle,|2\rangle=|\bar{3}\rangle$.
Here, the green squares and triangles   denote respectively the parameters of the $\Pi\Phi$
and $\Phi\Pi$ from Fig.~\ref{fig:inductance_qutrits}.(c) and (d).
Notice that, for a given $E_J$, the qutrit condition may be achieved for two different values of the flux bias.

\begin{figure*}[t]
    \centering
\includegraphics[width=0.258\linewidth]{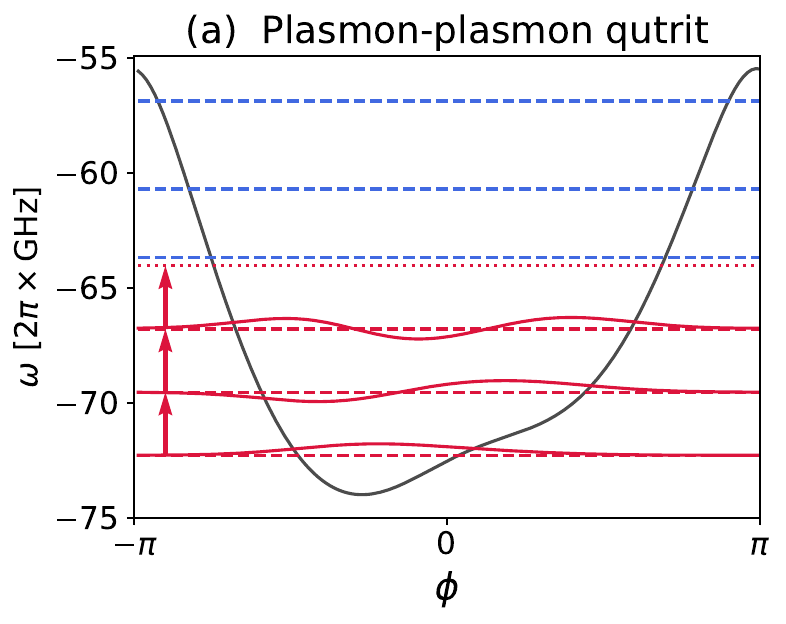}
\includegraphics[width=0.23\linewidth]{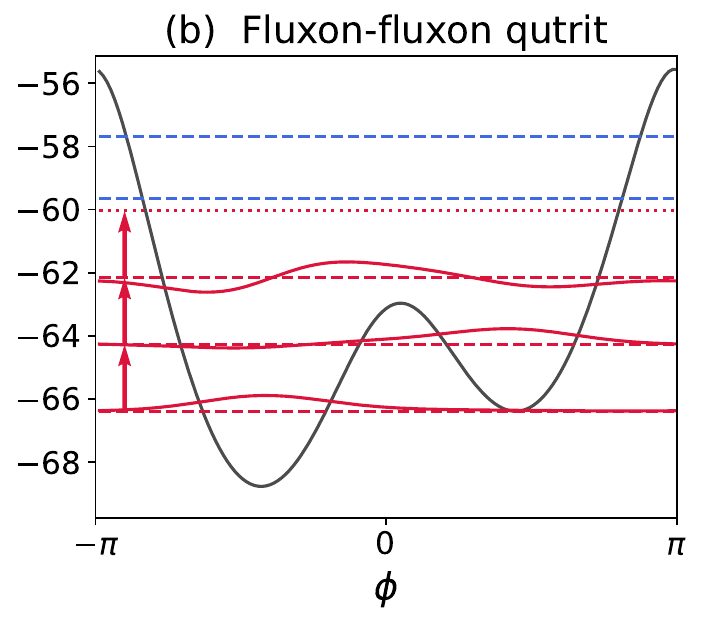}
\includegraphics[width=0.23\linewidth]{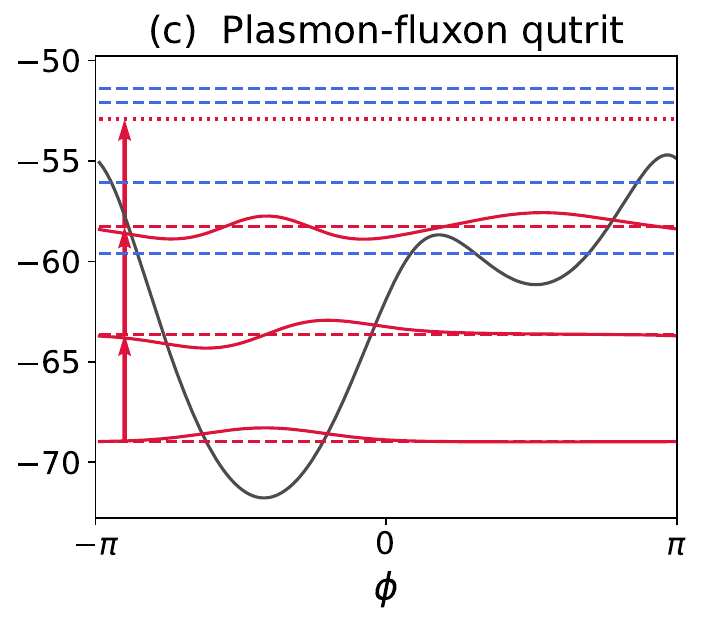}
\includegraphics[width=0.23\linewidth]{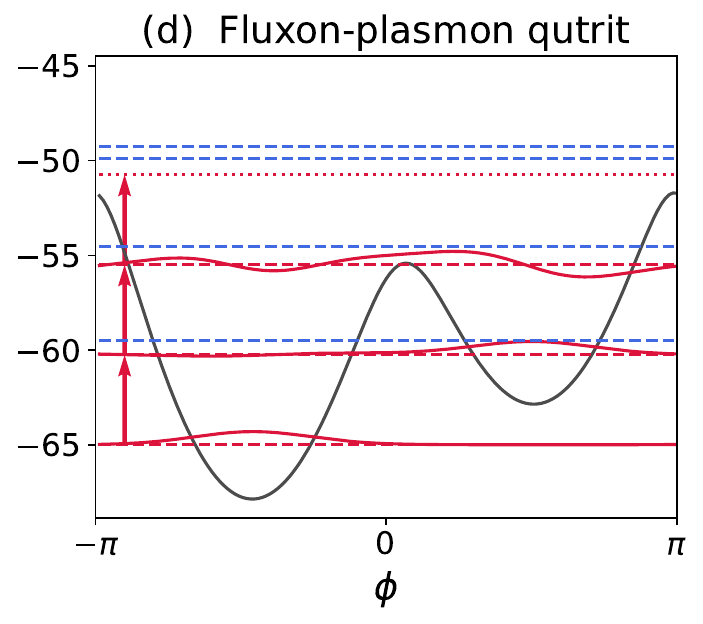}
    \caption{Qutrits arising in four different regimes of a circuit made with two higher harmoni JJs. The black
    solid line depicts the Josephson potential $V(\phi)$. 
    The red dashed
    lines represent the energy levels of the qutrit, and the solid red curves sketch the wavefunction of the states (not normalized). The blue dashed lines indicate the other levels
    of $H_{\rm at}$, which are off resonant with respect to the transitions at energy $\omega_{10}=\omega_{21}$ (the red dotted line providing a visual hint of the detuning).
    Panels (a-d) correspond to the plasmon-plasmon, fluxon-fluxon, plasmon-fluxon and fluxon-plasmon qutrits.
    }
    \label{fig:HHJJ_qutrits}
\end{figure*}

\section*{Appendix D: Qutrit regimes in higher harmonic JJ circuits}

In this Appendix, we will consider an alternative architecture to realize qutrit quantum simulators,
based on higher harmonic JJs.
We will show that similar level structures as in fluxonium can in principle be achieved; 
however,  the large  decoherence expected in practical implementations may pose a serious obstacle.

The search for 
 robust qubits  
and the study of Andreev bound states have stimulated
a recent outburst of interest towards engineering unconventional JJ with higher order nonlinearities~\cite{Larsen2015,deLange2015,Larsen2020,Messelot2024}.
When graphene, nanowires or semiconductors are used to implement the weak link between two superconducting islands, the energy-phase relation is not a simple cosine, but the Josephson potential takes the form~\cite{Beenakker1991,Glazman2021}:
\begin{equation}
    V_{\vec{T},\Tilde{\Delta}}(\phi) = - \Tilde{\Delta} \sum_j
    \sqrt{1- T_j \sin^2 \left( \frac{\phi}{2} \right)} \ ,
\end{equation}
where $\Tilde{\Delta}$ denotes the superconducting gap and $T_j$ the transmission of each Andreev channel across the junction. For a low transmission JJ with a single channel $T_0 \to 0$, one recovers the standard Josephson potential
$V \simeq -\Tilde{\Delta} (1-T_0/4) -\frac{\Tilde{\Delta} T_0}{4} \cos\phi \equiv - \Tilde{\Delta} + E_J - E_J \cos\phi$.

Two such JJs can be put in parallel and, by inserting a magnetic flux $\Phi$,  quantum interference leads to an effective junction described by the potential
\begin{equation}
    V(\phi) = V_{\vec{T}^A,\Tilde{\Delta}^A}(\phi) 
    +
    V_{\vec{T}^B,\Tilde{\Delta}^B}(\phi+\Phi) .
\end{equation}
In particular, for identical junctions ($\vec{T}^A=\vec{T}^B,\Tilde{\Delta}^A=\Tilde{\Delta}^B$) and maximally destructive interference $\Phi=\pi$,
only the even harmonics of $ V(\phi)$ are present (and two conventional JJs would just cancel out, $V \sim 0$).
This setting has been recently used to experimentally detect higher order Josephson harmonics in semiconductor and graphene junctions~\cite{Kringhoj2018,Larsen2020,Messelot2024}.

While for $\Phi=\pi$ the potential displays a symmetric double well structure, varying the external flux allows to achieve interesting asymmetric configurations. In particular,
one can recover regimes which are reminescent of  the $\Pi\Pi, \ \Phi\Phi, \ \Pi\Phi$ and $\Phi\Pi$ qutrits described above, without the need of any superinductance ($E_L=0$). 
For example, in the experimental work by Larsen et al.~\cite{Larsen2020}, they could vary the transmission coefficients $\vec{T}^A,\vec{T}^B$ by gating the InAs nanowires of the JJ. For different gate voltages, qubits operating either at a fluxon or plasmon transition were achieved, and they also measured the crossing of $\omega_{10}$ and $\omega_{21}$ by varying the external flux.

We  fix $\Delta^A=\Delta^B=
2\pi \times 45$~GHz (as in \cite{Larsen2020}) and consider junction $A$ to have a high transmission vector $\vec{T}^A=\{ 0.95 \}$ (we consider only one transmission channel for simplicity, but similar results are obtained when a few channels are present).
We vary the flux in the JJ loop $\Phi$ and 
the transmission of the $B$ junction
$\vec{T}^B=\{ T_B \}$. In practice, this can be done by gating the semiconductor or graphene or nanowires making up the JJ.
For example, in Ref.~\cite{Larsen2020}, for different gate voltages the transmission of the InAs nanowires could result in
 $ \vec{T}^A = \{1.0, 0.91, 0.30, 0.20, 0.18\}, \
\vec{T}^B = \{0.90, 0.06, 0.06, 0.06\}$ or
 $ \vec{T}^A = \{1.0,1.0,0.6\}, \
\vec{T}^B = \{0.99,0.78,0.31,0.3\}$.
In that same paper, the plasmon and fluxon regimes  were identified at the qubit level, and the condition $\Delta=0$ was also observed. 

The main goal of this Appendix is to show, via the four examples in Fig.~\ref{fig:HHJJ_qutrits}, that the four qutrit regimes can be achieved also with this kind of circuits.

In panel (a), we report the 
plasmon-plasmon regime, obtained for 
$T_B=0.65$ and $\Phi/\Phi_0=0.471$.
The qutrit  operates at a frequency of $\omega_{10} = 2\pi \times 2.75$~GHz
and is protected by a detuning of 
$\delta = 2\pi \times 374$~MHz.
Similarly to the results obtained for fluxonium circuits, we find 
$\alpha=1.1$ and $P/J=0.17$
for capacitative coupling.

Panel (b) contains the fluxon-fluxon regime, with parameters
$T_B=0.85$, $\Phi/\Phi_0=0.486$,  $\omega_{10} = 2\pi \times 2.12$~GHz, 
$\delta = 2\pi \times 399$~MHz, 
$\alpha=1.77$ and $P/J=9.54$.

Panel (c) contains the plasmon-fluxon regime, with parameters
$T_B=0.9$, $\Phi/\Phi_0=0.441$,  $\omega_{10} = 2\pi \times 5.36$~GHz, 
$\delta = 2\pi \times 829$~MHz, 
$\alpha=0.67$ and $P/J=9\cdot 10^{-4}$.

Panel (d) contains the fluxon-plasmon regime, with parameters
$T_B=0.98$, $\Phi/\Phi_0=0.474$,  $\omega_{10} = 2\pi \times 4.75$~GHz, 
$\delta = 2\pi \times 827$~MHz, 
$\alpha=3.73$ and $P/J=0.83$.

\begin{figure*}[t]
    \centering
\includegraphics[width=0.49\linewidth]{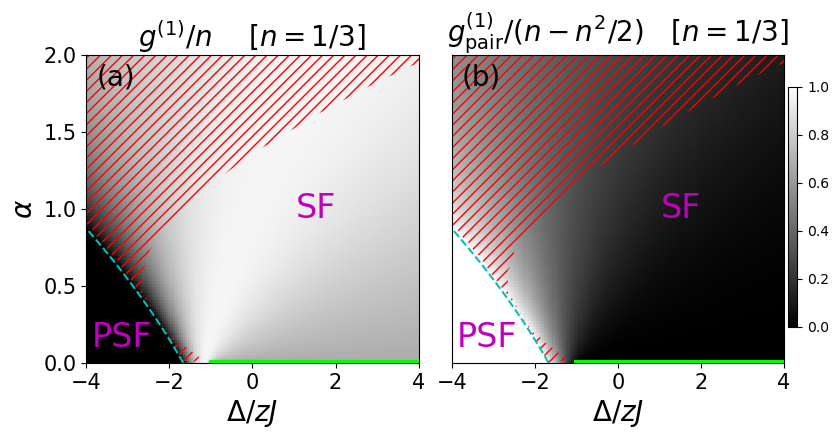}
\includegraphics[width=0.49\linewidth]{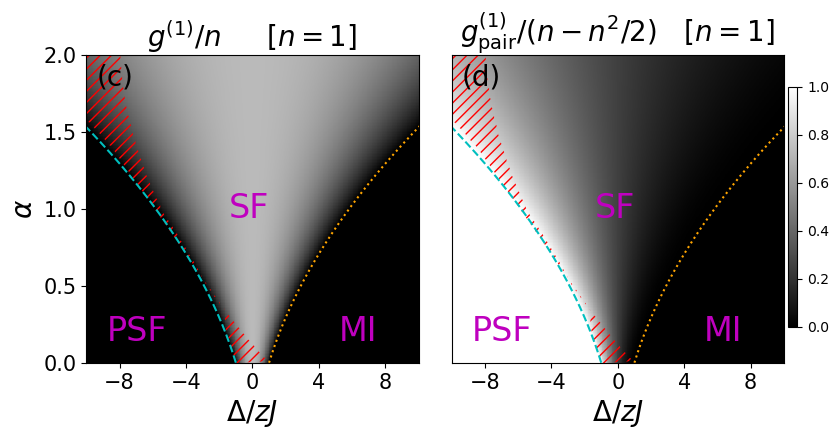}
    \caption{(a) Single-particle coherence $g^{(1)}=
\langle b^\dagger_i \ b_j \rangle$
of the ground-state of $H_{\alpha\Delta}$ for a small density $n=1/3$, calculated within the Gutzwiller approximation. (b) Pair coherence $g_{\rm pair}^{(1)}=
\langle (b^\dagger_i)^2 \ b_j^2 \rangle$.
In both panels,  
we can distinguish a standard superfluid (SF) and a (heavy-)pair superfluid (PSF$^*$) phase, delimited by cyan dashed line defined in Eq. (\ref{eq:DeltaPC}).
The red hatched area 
is defined by 
the thermodynamic instability condition
$\frac{d^2 e}{dn^2} <0$.
Finally, the green line at $\alpha=0$  and $\Delta \geq \Delta_*$  indicates the complete suppression of pairs, $\psi_2=0$.
Panels (c) and (d) are the analogous of (a) and (b),
but for a system at unit filling and $\Delta=2.5zJ$. In this case, a Mott insulator phase (MI) with one particle per site is present, delimited by the orange dotted line, calculated in Eq. (\ref{eq:DeltaMott}).
}
\label{fig:Gutzwiller_alfaDelta_App}
\end{figure*}

Importantly, we remark that, in the absence of the superinductance, the Josephson potential is periodic and the phase variable is defined modulo $2\pi$. This also entails that the charge bias $n_g$ cannot be gauged away, and results in a quite sizable sensitivity to charge fluctuations, with bandwidths~\footnote{The charge bias $n_g$ can be seen as a Bloch quasi-momentum, and correspondingly the energy levels form bands~\cite{Koch2007}.} of the order of hundreds of MHz for the values considered here. Such a  large charge-bias decoherence would make the qutrit impossible to use. 
Moreover, it seems unpractical to reduce the charge sensitivity by making the Josephson potential steeper (like in transmons), since, when  working at the large flux biases required for qutrits, $V(\phi)$ is the result of a destructive interference  and a realistic superconducting gap needs to be considered.   
One could instead think of combining
superinductances and higher-harmonic JJs.
Still, it is believed that the
gating of the JJs introduces additional noise channels (such as in the critical current and transmission coefficients~\cite{Pellegrino2020}).
Thus, at the state of the art,  these higher-harmonic JJ circuits are plagued by a large decoherence and 
qutrit realizations based on fluxonium currently seem much more promising. 
Nonetheless, research on unconventional JJs is a very active field and significant ground-breaking improvements may come out in the future.

\section*{Appendix E: Gutzwiller calculations}

In this Appendix, we report additional details concerning the Gutzwiller results presented in Sec. \ref{sec_many-body}.
We assume the array to consist in a bipartite lattice with coordination number $z$ and to have $J,P \geq 0$, excluding geometric frustration.

Assuming for starters translationally invariant states, the Gutzwiller approximation for the ground state $|\Psi\rangle$ of the array of qutrits reads
\begin{equation}
    |\Psi\rangle = \bigotimes_j \  (\psi_0 |0\rangle_j + \psi_1 |1\rangle_j+ \psi_2 |2\rangle_j ),
\end{equation}
where the $\{\psi_a\}_{a=0,1,2}$ specify the local wavefunction of each qutrit.
It is easy to see from the functional of the energy that for the ground state one can always choose the $\psi_n$ to be real and positive.
Fixing the normalization of the state $\sum_n \psi_n^2 = 1$ 
and the density $\psi_1^2 + 2\psi_2^2 = n$ allows to parametrize the ground state via the only parameter $\psi_2 \in [\sqrt{\max\{0,n-1\}}, \sqrt{n/2}]$, according to
\begin{equation}
    \psi_0 = \sqrt{1-n+\psi_2^2} \ \ , \ \ \ \psi_1 = \sqrt{n-2\psi_2^2}.
\end{equation}
The energy functional per site then reads
\begin{multline}
   E[\psi_2] = 
   \langle \Psi | \hat{H} | \Psi \rangle / L = 
   -zJ (\psi_0 \psi_1 + \sqrt{2} \alpha \psi_1 \psi_2)^2 \\ - zP(\psi_0 \psi_2)^2 + \Delta \psi_2^2
   +
  \frac{z}{2} (W_0\psi_0^2 + W_1\psi_1^2 + W_2\psi_2^2)^2
   ,
\end{multline}
where $z=2d$ for a square lattice in $d$ dimensions.

The energy can be minimized to find the optimal $\psi_2$. A single-particle condensate is signaled by a finite value of the off-diagonal coherence
$g^{(1)}=
\langle b^\dagger_i b_j \rangle
=
(\psi_0\psi_1 + \sqrt{2}\psi_1\psi_2)^2$, with $i \neq j$. Similarly, condensation of pairs corresponds to $g_{\rm pair}^{(1)}=
\langle (b^\dagger_i)^2 \ b_j^2 \rangle
= 2(\psi_0\psi_2)^2$. 
At density $n=1$, a Mott insulating state can occur with $\psi_1=1, \psi_0=\psi_2=0$. 
Moreover, the thermodynamic stability 
of the state is assessed by looking at $\frac{d^2 e}{dn^2}$. When this is negative, it signals instabilities towards forming high density clusters.

\begin{figure*}[t]
    \centering
\includegraphics[width=0.49\linewidth]{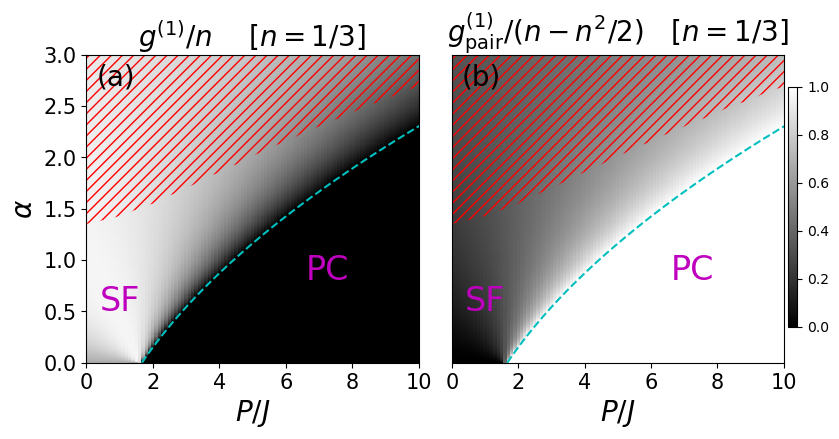}
\includegraphics[width=0.49\linewidth]{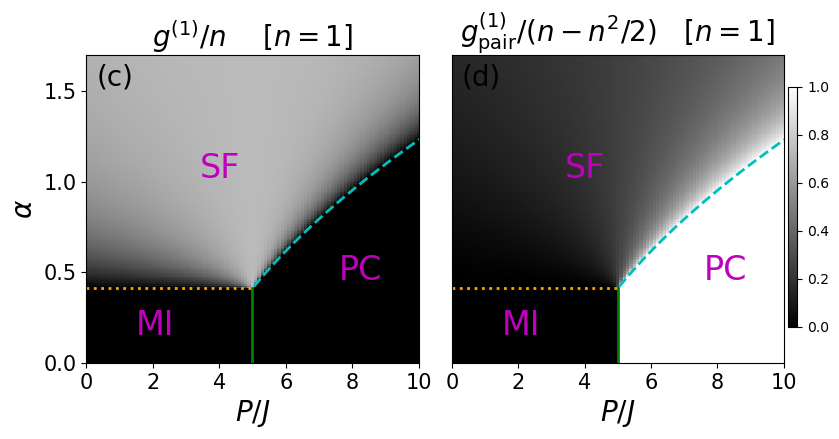}
    \caption{(a) Single-particle coherence $g^{(1)}=
\langle b^\dagger_i \ b_j \rangle$
of the ground-state of $\hat{H}$ as a function
of the hopping correlation $\alpha$ and of the pair-hopping strength $P/J$. Here, we take  $\Delta=0$
and  a small density $n=1/3$, and perform the calculations within the Gutzwiller approximation. (b) Pair coherence $g_{\rm pair}^{(1)}=
\langle (b^\dagger_i)^2 \ b_j^2 \rangle$.
In both panels,  
we can distinguish a standard superfluid (SF) and a pair superfluid (PSF) phase, delimited by the 
cyan dashed line defined by Eq. (\ref{eq:P_SFPSF}).
The red hatched area 
is defined by 
the thermodynamic instability condition
$\frac{d^2 e}{dn^2} <0$.
Panels (c) and (d) are the analogous of (a) and (b), with the difference that now we take unit filling $n=1$ and a finite repulsive Hubbard interaction $\Delta=2.5 zJ$.
At this filling, we also have a Mott insulator phase (MI) with one particle per site, delimited by the orange horizontal dots and the green vertical line, defined in the text.
}
\label{fig:Gutzwiller_alfaP}
\end{figure*}

\subsubsection*{Phase diagram in the  $(\alpha,\Delta)$ plane}

We start by considering the case  $P,W_r=0$, determined by the competition between correlated hopping and local interactions.
In fact, when $\Delta/J \ll -1$, it is favorable to populate the state
$|2\rangle$, resulting in the formation of as many pairs as possible, i.e. $\psi^2_2 = n/2$.
This is the PSF$^*$ regime; we use the asterisk to recall that, in this regime, the superfluid stiffness is expected to be small, of the order of $J^2\alpha^2/|\Delta|$.
The boundary of the  PSF$^*$ phase is found expanding around this value, to obtain the condition
\begin{equation}
    \Delta = -\frac{2zJ}{n}
    \left(
    \sqrt{n - n^2/2} + n \alpha
    \right)^2.
    \label{eq:DeltaPC}
\end{equation}
The boundary of this PSF phase corresponds to the cyan dashed line in Fig.~\ref{fig:Gutzwiller_alfaDelta_App}. 
Instead, when $\alpha=0$ and  at small $\psi_2$,  one has $e[\psi_2] \simeq -zJ(n-n^2) + [\Delta-zJ(3n-2)]  \psi_2^2 + ...$. This entails $\psi_2=0$ for $\Delta>\Delta_*=2J(3n-2)$, and physically corresponds to a Bose gas of perfectly impenetrable bosons. For small $\alpha$, $\psi_2 \propto \alpha$.

The Gutzwiller results 
for $g^{(1)}$ and $g_{\rm pair}^{(1)}$
at a small density 
$n=1/3$
are reported in Fig.~\ref{fig:Gutzwiller_alfaDelta_App}.(a) and (b), respectively. 
At this density, the SF and PSF$^*$ are the only two thermodynamically stable phases. 
The red hatched area represents the thermodynamically unstable region discussed in the main text and in the Appendix F.

Having analyzed the small density phase diagram of the model, we would like to recall the filling $n=1$ case, already discussed in the main text.
The corresponding single-particle and pair coherences are reported in Fig.~\ref{fig:Gutzwiller_alfaDelta_App}.
The Mott insulator (MI) state 
$\psi_1=1,\psi_0=\psi_2=0$
is now the favorable phase for
\begin{equation}
    \Delta
    \geq
    zJ (1+\sqrt{2}\alpha)^2,
    \label{eq:DeltaMott}
\end{equation}
obtained expanding the energy around $\psi_2=0$ and  indicated by the orange dotted line in the plots.

\subsubsection*{Phase diagram in the  $(\alpha,P)$ plane}

We now assume $P \neq 0$.
The case of negligible Hubbard interactions $\Delta=0$ and small density $n=1/3$ is reported in Fig.~\ref{fig:Gutzwiller_alfaP}.(a) and (b),  for $g^{(1)}$ and $g^{(1)}_{\rm pair}$, respectively.
The scenario with unit filling $n=1$
and $\Delta=2.5zJ$ is instead illustrated in panels (c) and (d).

The PSF is stabilized for
\begin{equation}
P \geq
\frac{2J}{n}
\left(
\sqrt{n-n^2/2} + n\alpha
\right)^2 + z^{-1}\Delta,    
\label{eq:P_SFPSF}
\end{equation}
represented as the cyan dashed line.
For $n=1$, the MI occurs in the rectangle defined by
$\alpha \leq \sqrt{\frac{\Delta}{2zJ}} - \frac{1}{\sqrt{2}}$, indicated by the horizontal orange dots,
and  $P \leq 2z^{-1}\Delta$, given by the vertical green line.

\subsubsection*{Impact of the $\hat{W}$ term.}

When  nonlocal interactions $W_r \neq 0$ are present, the system can undergo spontaneaous symmetry breaking to crystalline phases.
 For a bipartite lattice (e.g. a 1D chain or a 2D square lattice) with sublattices $C$ and $D$,
the Gutzwiller ansatz  takes the form
\begin{equation}
    |\Psi\rangle = 
    \bigotimes_{k \in C}  \
      \left( \sum_{r=0}^2 c_r |r\rangle_k
      \right)
    \bigotimes_{l \in D}  
    \  \left( \sum_{r=0}^2 d_r |r\rangle_l
      \right),
\end{equation}
Introducing the density imbalance order parameter $m$,
the density on $C$ and $D$ sites is given by
$n+m$ and $n-m$, respectively.
Keeping into account of the normalization of the wavefunctions and noticing that if $J \cdot P>0$ we can take the wavefunction coefficients to be real,
we obtain
$c_0 = \sqrt{1-n-m+c_2^2}$ and $c_1 = \sqrt{n+m-2c_2^2}$
for
$c_2 \in [\sqrt{\max\{0,n+m-1\}}, \sqrt{(n+m)/2}]$, and similarly for the $d_r$'s but replacing $m \to -m$.

The Gutzwiller energy per site reads
\begin{multline}
    \epsilon[c_2,d_2,m]
    =
    -zJ \langle \alpha^{\rho}b \rangle_C  \langle \alpha^{\rho}b \rangle_D -zP c_0c_2d_0d_2
    \\
    + \frac{\Delta}{2}(c_2^2 + d_2^2)
+ \frac{z}{2} {\rm sign}(g) 
\langle \hat{W}(\hat{\rho)}\rangle_C
\langle \hat{W}(\hat{\rho)}\rangle_D,
\end{multline}
with 
$\langle \alpha^{\rho}b \rangle_C = c_0 c_1 + \sqrt{2} \alpha c_1 c_2$
and 
$\langle \hat{W}(\hat{\rho)}\rangle_C = 
W_0 + \delta W_1 c_1^2 +  \delta W_2 c_2^2
$, and similarly for the $D$ sublattice.
The energy can be optimized numerically as a function of $c_2, d_2, m$ and fairly complex behaviors can be found.

\begin{figure*}
    \centering
\includegraphics[width=0.98\linewidth]{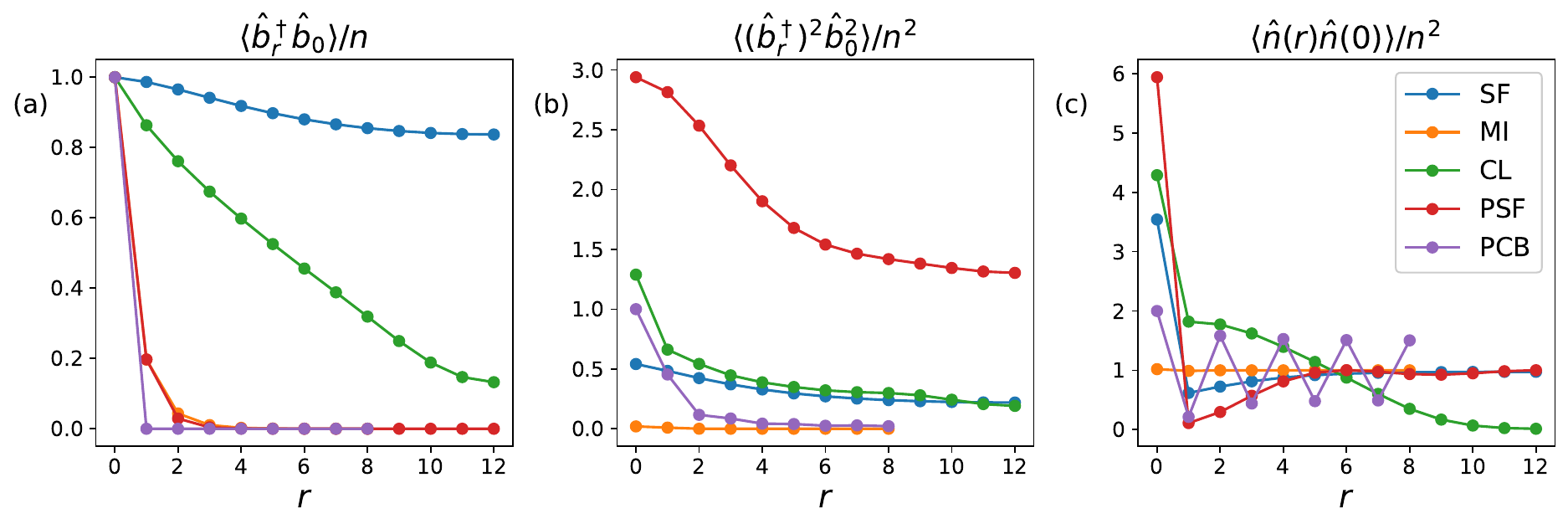}
    \caption{ Ground-state correlation functions obtained from exact diagonalization. Different regimes --labeled SF in blue, MI in orange, CL in green, PSF in red and PCB in violet-- of the qutrit Hamiltonian of Eq. (\ref{eq:the_H})
    are reported, as described in the text.
    Panel (a) displays, as a function of distance $r$, the single particle coherence function
    $g^{(1)}(r)$,
    the pair coherences $g^{(1)}_{\rm pair}(r)$
    are obtained in  (b), and panel (c) reports
    the density-density correlator.
}
\label{fig:correls}
\end{figure*}

Here, we focus on the simple case $J=0$ and $n=1$, where single-particle hopping processes can be disregarded and simple analytical results are obtained.
The MI has then energy
$\epsilon_{\rm MI} = \frac{z}{2} \delta W_1^2$, while
$\epsilon_{\rm PSF} = 
-zP (1-\frac{n}{2})\frac{n}{2}
+
\frac{\Delta}{2}n
+
\frac{z}{8} \delta W_2^2 n^2$
for the PSF.
Breaking translational symmetry ($m\neq0$) leads to a
pair checkerboard (PCB) phase with
wavefunction
$c^2_2=(n+m)/2$, $d^2_2=(n-m)/2$ and $c_1=d_1=0$.
The energy reads
$\epsilon_{\rm PCB}
=
-zP \sqrt{
\frac{(2-n)^2-m^2}{4}}
\sqrt{\frac{n^2-m^2}{4}
}
+
\frac{\Delta}{2}n
+
\frac{z}{8} \delta W_2^2 (n^2-m^2)
$,
which reduces to $\epsilon_{\rm PSF}$
for $m=0$.
Since for $n=1$ we have
$\epsilon_{\rm PCB}
=
\frac{z}{8}
\left[ 
-2P
+ \delta W_2^2
\right] (1-m^2)
+
\frac{\Delta}{2}
$,
there are only two possibilities, $m=0$ and $m=1$, with the PSF-PCB transition occurring at
$P=
 \delta W_2^2 / 2$.
Also, notice that pair hopping is completely suppressed in the PCB phase.
The energy of the broken symmetry phase is then simply
$\epsilon_{\rm PCB}
=\frac{\Delta}{2}$.

The PCB to MI transition
occurs instead for 
$\Delta=z \delta W_1^2$, while
the PSF to MI transition occurs for
$\Delta = \frac{z}{2}P \left( 1 - \frac{\delta W_2^2}{2zP}
\right) + z \delta W_1^2$.
These results are summarized in Fig.~\ref{fig:Gutzwiller_W_impact} of the main text.

\section*{Appendix F: Numerical ground state results}

In this Appendix, we use exact diagonalization to validate our Gutzwiller analysis for the ground state of the qutrit Hamiltonian $\hat{H}$.

\subsubsection*{Correlation functions}

Since within exact diagonalization symmetries cannot be spontaneously broken,
one cannot classify ground states based on order parameters. but needs to study the behavior of the correlations functions.
We do this in Fig.~\ref{fig:correls}, by picking illustrative sets of parameters from the different phases supported by $\hat{H}.$

For each regime, we plot in panel (a) the corresponding single particle coherence function
$g^{(1)}(r) = \langle \hat{b}^\dagger_r \hat{b}_0 \rangle$, as a function of spatial distance $r$ and at equal times. In panel (b) we instead report the pair coherences
$g^{(1)}_{\rm pair}(r) = \langle (\hat{b}^\dagger_r)^2 \hat{b}_0^2 \rangle$,
while the density-density correlator
$\langle \hat{n}(r) \hat{n}(0) \rangle$ is displayed in (c).

\begin{figure*}
    \centering
\includegraphics[width=0.98\linewidth]{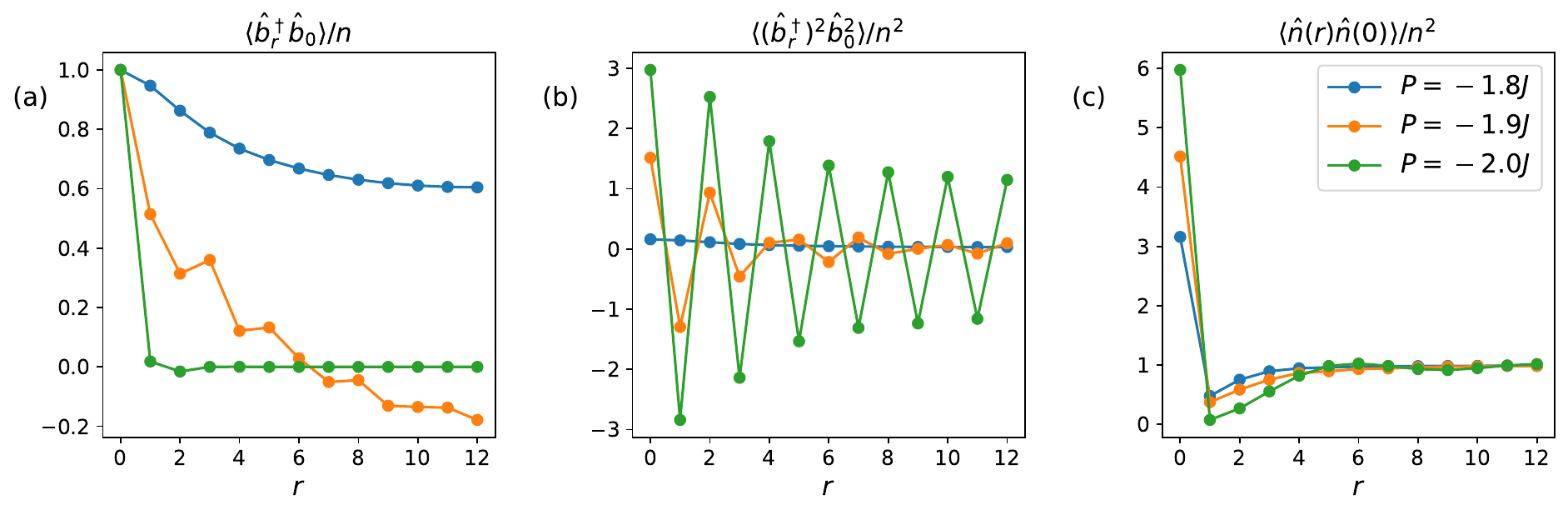}
    \caption{ Ground-state correlation functions obtained from exact diagonalization.
    The competition between $J>0$ and $P<0$ is studied here, for 
    $L=24, N=8, \alpha=1,\Delta=w_r=0$. 
    Panel (a) displays, as a function of distance $r$, the single particle coherence function
    $g^{(1)}(r)$,
    the pair coherences $g^{(1)}_{\rm pair}(r)$
    are obtained in  (b), and panel (c) reports
    the density-density correlator.
}
\label{fig:correls_TSF}
\end{figure*}

The first phase, corresponding to the blue numerical points,
is the standard superfluid. This is obtained here for $\alpha=1$ and
$P,\Delta,W_r=0$ within a one-dimensional array of length $L=24$ and containing $N=8$ photons ($n=1/3$ filling).
The SF is characterized by a large $g^{(1)}(r)$
and a sizable $g^{(1)}_{\rm pair}(r)$,
while the density-density correlator displays bunching at $r=0$ and a slight anti-bunching at finite $r$, due to the three-body constraint.

A Mott insulator is obtained for $\Delta=20$
and $n=1$ filling, for
$L=N=16$
(orange points).
In this case, the $g^{(1)}(r)$ decays very rapidly and  $g^{(1)}_{\rm pair}(r)$ is negligible, while 
$\langle \hat{n}(r) \hat{n}(0) \rangle \simeq 1 \ \forall r$, corresponding to a ground state very close to the product state with one particle per site.

The clustered state (CL) is represented by the green data. This is obtained at $n=1/3$ filling
($L=24,N=8$) and is driven by the correlated hopping term with $\alpha=1.5$.
The CL ground state can be seen as the quantum superposition of a collapsed quantum droplet
centered at different sites (so to preserve translational invariance). In the thermodynamic limit, the droplet would be stabilized by the three-body constraint.
This physics results in the decay of all correlations functions at large distance (since there are no particles outside of the droplet).
This is particularly striking in the density-density correlator, 
since in any stable and translational invariant state $\langle \hat{n}(r) \hat{n}(0) \rangle$
should tend to $n$ for large $r$.

The red points illustrate the pair superfluid state obtained for $P/J=10$ and $L=24,N=8$.
While the single particle coherence decays very quickly (similarly to the MI), large values and a slow decay are observed for $g^{(1)}_{\rm pair}(r)$.
The density correlations are dominated by the binding of the pairs at $r=0$ and by the hard-core repulsion between the pairs at small $r$'s.

Finally, the pair checkerboard (PCB) phase corresponds to the violet data.
This is obtained here for $J=0$ but finite $P$,
and unit filling $L=N=16$. It is driven by the competition between the pair hopping  and 
the repulsive nearest-neighbor interactions,
quantified by $w_1=0.5, w_2=2$.
Since the particles only come in pairs, 
$g^{(1)}(r)=0 \ \forall r\neq 0$.
Instead, $g^{(1)}_{\rm pair}(r)$ is finite but decays quickly, since the state is insulating.
However, the most distinctive signature of the PCB is seen in the staggered behavior of the density-density correlator, reflecting the crystalline order of having one pair every other site.

In conclusion, the correlators displayed in  Fig. \ref{fig:correls} provide an exact diagonalization confirmation of the results obtained using the Gutzwiller ansatz.

\subsubsection*{The $J>0,P<0$ case.}

As anticipated in the main text,
a kinetic term with $J>0$
favors a superfluid with uniform phase across the system,
while pair hopping with $P<0$ is optimized by a pair superfluid with pair momentum $k=\pi$.
While for the qutrits we found
that Gutzwiller predicts a sharp transition between these two phases,
a regime with strong fluctuations
is revealed by exact diagonalization.
In Fig.~\ref{fig:correls_TSF} we report the correlation functions for a closed chain of $L=24$ sites at $1/3$ filling $N=8$ and for four values of  $P/J$ (and $\alpha=1, \Delta=w_r=0$). 
As in the previous paragraph,
panel (a) displays the single particle coherence function over space, panel (b) the pair coherences, and (c) the density-density correlator.

For $P=-1.8J$ (blue points), the single particle hopping term prevails, leading to large single particle coherences. Notice that, due to the competition of the two kinetic processes, the pair coherences are suppressed compared to the $P=0$ case (shown in blue in Fig.~\ref{fig:correls}).
At $P=-2J$ (green points), instead, the pair coherences dominate and display a staggered pattern corresponding to pairs with $k=\pi$ momentum. 
The most interesting scenario occurs for $P=-1.9J$ (orange points), where both single-particle and pair coherences are important and display a complex pattern.
We interpret this frustrated ground state as the remnant of the twisted superfluid observed without 3-body constraint~\cite{jurgensen2015twisted,Luhmann}.

\subsubsection*{Ground state at small $\alpha$.}

\begin{figure*}[t]
    \centering
\includegraphics[width=0.98\linewidth]{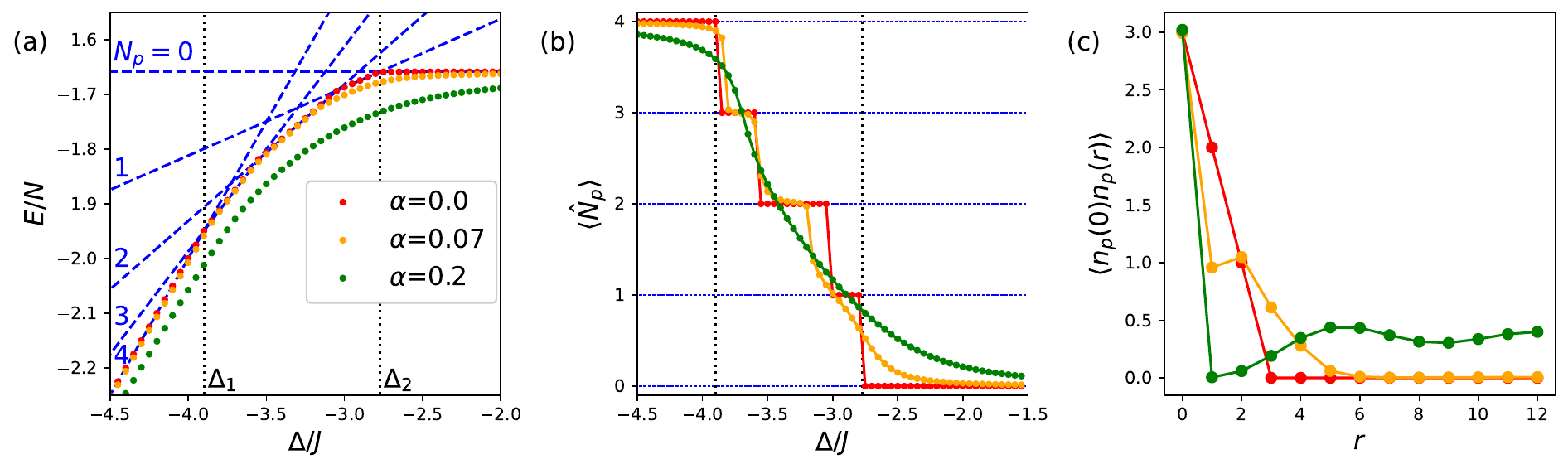}
    \caption{ Physics of the ground-state of $\hat{H}_{\alpha \Delta}$ at small $\alpha$, for a system of $N=8$ photons over $L=24$ sites. The red, yellow and green points correspond to $\alpha=0,0.07,0.2$, respectively.
    (a) Energy per particle
    as a function
    of $\Delta/J$. The blue dashed lines are the analytical prediction of Eq. (\ref{eq:exactE_small_alfa}), describing phase separation between a cluster of pairs and a Tonks-Girardeau gas of single particles. Each line is labeled by the quantum number $N_p$, being the number of pairs.
    As indicated by the vertical black dotted lines, for $\Delta>\Delta_2$ the ground state is Tonks-Girardeau gas with $N_p=0$, while for $\Delta<\Delta_1$ all particles are paired up (and can't move at $\alpha=0$, so that the ground state is degenerate).
     The number of pairs is better studied in panel (b).
     Finally, in panel (c) the pair-pair density correlation function is displayed.
}
\label{fig:small_alfa_trittic}
\end{figure*}

The physics occurring at small $\alpha$ deserves special attention.
We already saw, at the Gutzwiller level, that for very negative $\Delta$ all particles are paired ($\psi_2^2=n/2$), while at larger $\Delta$ one has a conventional superfluid of (nearly) impenetrable bosons, resulting in $\psi_2 = 0$ at $\alpha=0$. In between, an unstable region was predicted. 
Here, we study the 1D case, where exact solutions can be obtained for $\alpha=0$ and ED is used for finite $\alpha$. We also set $P, W_r=0$.

Let's start with the $\alpha=0$ case, for which the number of pairs
$\hat{N}_p = \sum_j (b_j^\dagger)^2 b_j^2$
is a conserved quantum number, since hopping can occur only from a singly occupied site to an empty one.
At large enough $\Delta$, it is unfavorable to have pairs and the ground state lives in the sector $N_p=0$.
The ground state is a Tonks-Girardeau~\cite{Tonks1936,Girardeau1960} lattice gas of impenetrable bosons,  with energy
\begin{equation}
    E_{TG}(N,L) = -2J\sum_{n=-\frac{N}{2}+1/2}^{\frac{N}{2}-1/2}
    \cos \left(
    \frac{2\pi}{L}n
    \right).
\end{equation}
Here and in the following we assume that $N$ is even.
Then, for sufficiently negative $\Delta$, on the contrary, all particles are paired up, $N_p=N/2$, and the kinetic energy is zero.
At intermediate $\Delta$, there can be $N_p$ pairs and it is easy to convince oneself that the lowest energy configuration is achieved by accumulating all pairs next to each other and leaving the remaining $L-N_p$ sites for the unpaired 
$N-2N_p$ bosons. 
These behave like impenetrable bosons confined in a box, 
and fermionization entails that the energy of $N$ such bosons on a chain of $L$ with open boundary conditions
is given by
\begin{equation}
    E_{well}(N,L)
    =
    -2J \sum_{n=1}^N \cos 
    \left(
\frac{\pi}{L+1} n.
    \right)
\end{equation}
To summarize, the energy of the ground state at $\alpha=0$ and in the 
sector with $N_p$ pairs
reads
\begin{multline}
    E_{\alpha=0}(N,N_p,L)
    =
    \\
    \begin{cases}
    E_{TG}(N,L), & \text{if } N_p=0; \\
    \Delta N_p + E_{well}(N-2N_p,L-N_p), & \text{otherwise.}
    \end{cases}
    \label{eq:exactE_small_alfa}
\end{multline}
The energies 
$E_{\alpha=0}(N,N_p,L)$
 are plotted in Fig.~\ref{fig:small_alfa_trittic}.(a) as blue dashed lines for $N=8,L=24$, with the number in blue indicating the corresponding $N_p$ sector. 
For $\Delta \in (\Delta_1, \Delta_2)$ the ground state 
possess $0<N_p<N/2$ pairs and
displays the aforementioned phase separation between a cluster of pairs and a Tonks-Girardau gas. 
For each $N$ and $L$, one can obtain $\Delta_1$
by requiring 
$E_{\alpha=0}(N,N/2,L) = E_{\alpha=0}(N,N/2-1,L)$,
and $\Delta_2$ from $E_{\alpha=0}(N,0,L) = E_{\alpha=0}(N,1,L)$.
This interval is indicated in Fig.~\ref{fig:small_alfa_trittic}.(a) and (b) by the vertical black dotted lines.
In the thermodynamic limit $L \to \infty$ 
at fixed $N/L$, we find
$\Delta_1 \to -4 J$ and $\Delta_2 \to -J \left[ 1+3\cos
\left(
\frac{\pi N}{L} 
\right) \right]
$.

These analytical insights are confirmed by ED and 
explain the behavior of the system at finite $\alpha$.
In Fig.~\ref{fig:small_alfa_trittic}.(a), we plot
the energy of $N=8$ photons on a ring of $L=24$ sites 
as a function of $\Delta/J$ and for three different
values of 
$\alpha=0, \ 0.07, \ 0.2$, corresponding to the red, orange and green points, respectively.
The $\alpha=0$ result perfectly follows the lowest of the blue dashed lines obtained from Eq.~(\ref{eq:exactE_small_alfa}). 

In Fig.~\ref{fig:small_alfa_trittic}.(b), we report the expectation value of $\hat{N}_p$ on the ground state of the system. At $\alpha=0$, the number of pairs jumps between $N/2$ ad $0$ with integer steps, occurring where the $E_{\alpha=0}(N,N_p,L)$ energies cross each other.
Increasing $\alpha$ smoothens out this staircase-like curve.

\begin{figure*}
    \centering
\includegraphics[width=0.99\linewidth]{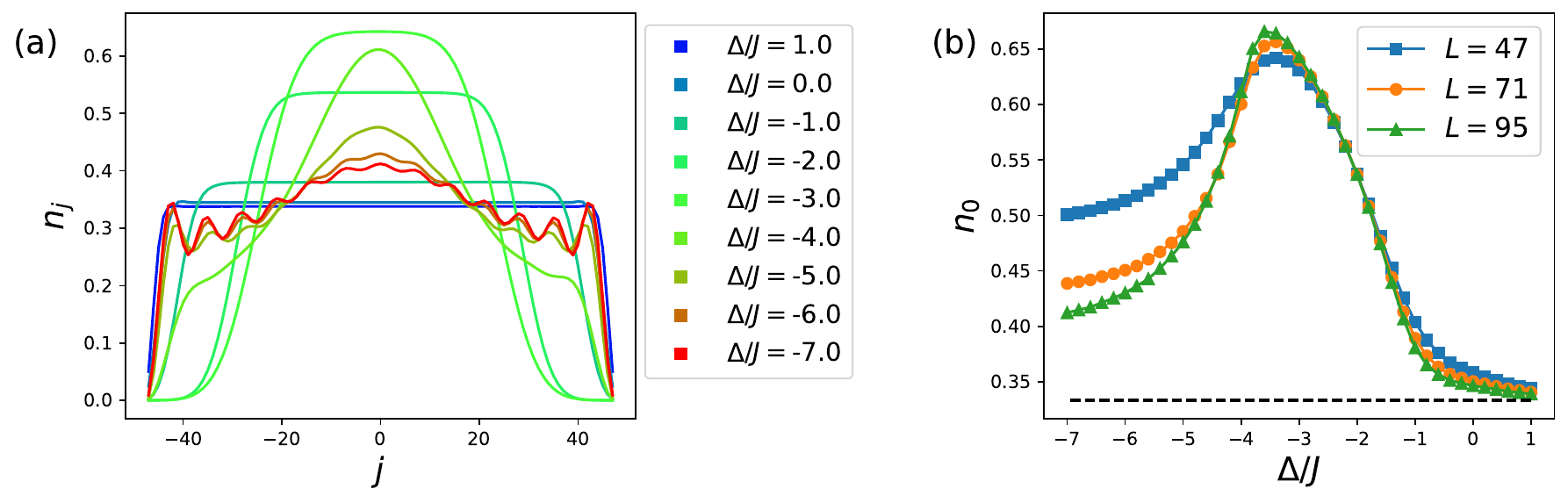}
    \caption{(a) Density of 
    a Bose-Hubbard model with three-body constraint for different values of the interaction $\Delta/J$.
    The results are obtained through a DMRG calculation 
    with $L=95,N=31$ and open boundary conditions.
    (b) We focus on the central site of the chain, for three different system sizes. The cluster instability is signalled by the peak at intermediate $\Delta/J$, becoming more prominent with $L$.
    }
\label{fig:DMRG}
\end{figure*}

Direct evidence that the pairs form a compact block, positioning themselves close to each other, is reported in Fig.~\ref{fig:small_alfa_trittic}.(c).
Here, we plot the pair density-density correlation function, defined as 
$\langle \hat{n}_p(0) \hat{n}_p(r) \rangle
=
\frac{1}{L} \sum_j \langle \hat{n}_p(j) \hat{n}_p(j+r) \rangle$, with 
$
\hat{n}_p(j) = \frac{1}{2} (\hat{b}^\dagger_j)^2 \hat{b}_j^2
$
and $r$ the distance.
In this particular case, we fix $\Delta/J=-3.7$,
for which we have $N_p=3$ at $\alpha=0$.
Since the three-body constraint entails $\hat{n}^2_p(j) = \hat{n}_p(j)$, we have that $L\langle \hat{n}_p(0) \hat{n}_p(0) \rangle = \langle \hat{N}_p \rangle$, confirming our expectation.
Moreover, the fact that $L\langle \hat{n}_p(0) \hat{n}_p(1) \rangle = 2$, $L\langle \hat{n}_p(0) \hat{n}_p(2) \rangle = 1$ and $L\langle \hat{n}_p(0) \hat{n}_p(r) \rangle = 0$ for $|r|>2$ is consistent with the three pairs being next to each other. 
The block of pairs fluctuates but persists for $\alpha=0.07$, while it disappears at $\alpha=0.2$.
The phase separation is also confirmed by keeping track of the stability condition $E(N+2) + E(N-2) - 2 E(N)>0$, which is indeed violated for $\alpha=0, \ 0.07$ at  
$\Delta/J=-3.7$ (not shown). The point $\alpha=0.2$ appears to be stable, instead, compatibly with 
$\langle \hat{n}_p(0) \hat{n}_p(r) \rangle$
tending to a plateau at large $r$.

\subsubsection*{Stability at $\alpha=1$}

When $\alpha=1$ and $\Delta<0$, our system resembles a standard Hubbard model with attractive interactions. While in the standard case the system would just collapse for any small $\Delta<0$, the three body constraint stabilize the system at large negative $\Delta/J$. In this regime, the ground state consists of a Tonks-Girardeau gas of pairs with mass $\propto \Delta/J^2$.
However, we found, consistently with the Gutzwiller predictions, that a collapse instability can indeed occur for  a window of intermediate values around $\Delta/J \sim 3$.
This was obtained performing density-matrix renormalization group (DMRG) simulations~\cite{white1992}
in a 1D chain with open boundary conditions. The ITensor library was used~\cite{itensor}.
The results are shown in Fig. \ref{fig:DMRG}.
In panel (a) we plot the density of a $L=95,N=31$ system (filling $n \simeq 1/3$) for different values of $\Delta/J$ (colder to warmer colors correspond to more negative values).
While for $\Delta/J \sim 0$ the physics is that of a weakly interacting gas confined in a box, with a characteristic healing length at the edges, for $\Delta/J =7$ we observe Friedel oscillations typical of a fermionized Tonks-Girardeau gas of pairs. The oscillation period is determined by the density, via the Fermi momentum.
For intermediate $\Delta/J$, the density self-focuses at the center of the box.
The density at the center of the box $n_0$ is plotted in Fig. \ref{fig:DMRG}.(b) as a function of $\Delta/J$. As noted above, $n_0$ peaks for intermediate values of $\Delta/J$. Moreover, this peak becomes more pronounced for larger system sizes, suggesting that this is not a finite-size effect.
On the contrary, outside of the unstable region, $n_0$ approaches the bulk density $n=1/3$ faster for larger $L$.
A precise characterization of the boundaries of this instability goes beyond the scope of this work.
Finally, while this calculation was performed at one third filling,
we remark that no instability is expected at integer density~\cite{Cuzzuol2025}.

\bibliography{bibliography}

\end{document}